\documentclass[aps,prx,reprint,footinbib,longbibliography]{revtex4-2}

\usepackage{graphicx}
\usepackage{bm,bbm}
\usepackage{amsmath}
\usepackage{amssymb}
\usepackage{mathtools}
\usepackage{siunitx}
\usepackage{makecell}
\usepackage{physics}
\usepackage[normalem]{ulem}
\usepackage{ragged2e}
\usepackage[table,dvipsnames,svgnames]{xcolor}
\definecolor{webblue}{HTML}{2D3092}
\usepackage[colorlinks, urlcolor=webblue, citecolor=webblue, linkcolor=webblue]{hyperref}
\usepackage[capitalize]{cleveref}

\usepackage{etoolbox}

\providecommand{\bibinfo}[2]{#2}
\renewcommand{\bibinfo}[2]{%
  \ifstrequal{#1}{title}{\emph{#2}}{#2}%
}

\usepackage{comment}
\usepackage{array}
\usepackage{multirow}
\usepackage{braket}
\usepackage{slashed}
\usepackage[english]{babel}
\usepackage[autostyle, english = american]{csquotes}

\usepackage[scr=kp]{mathalpha}

\makeatletter
\AtBeginDocument{%
  \renewcommand{\NAT@open}{\textcolor{webblue}{[}}%
  \renewcommand{\NAT@close}{\textcolor{webblue}{]}}%
}
\makeatother

\let\originaleqref\eqref
\renewcommand{\eqref}[1]{%
  \textcolor{webblue}{\originaleqref{#1}}%
}

\addto\captionsenglish{%
}

\begin{document}

\title{Dual gauge theory for two-dimensional superfluid turbulence}

\author{Tobias Helbig} 
\email{helbig@stanford.edu}
\affiliation{Leinweber Institute for Theoretical Physics, Stanford University, Stanford, CA 94305, USA}

\author{Sayak Bhattacharjee}
\email{sayakbhattacharjee@stanford.edu}
\affiliation{Leinweber Institute for Theoretical Physics, Stanford University, Stanford, CA 94305, USA}

\author{Srinivas Raghu}
\email{sraghu@stanford.edu}
\affiliation{Leinweber Institute for Theoretical Physics, Stanford University, Stanford, CA 94305, USA}

\date{\today}
\begin{abstract}
We describe turbulent hydrodynamics of superfluids in two spatial dimensions through a dual formulation, in which point-like vortices are coupled to an emergent 2+1 dimensional $U(1)$ gauge field. The turbulent cascade of superfluid kinetic energy manifests as transfer between modes of the dual electric field. The dual equations of motion directly yield superfluid hydrodynamics. We numerically evolve these in the presence of forcing and dissipation and find that the system approaches a turbulent statistical steady state. In the nearly incompressible regime, we observe clustering of like-signed vortices and an approximately constant negative incompressible energy flux over a broad wavenumber range, signalling the presence of an inverse energy cascade. The incompressible kinetic energy spectrum is consistent with Kolmogorov's scaling law for two-dimensional fluid turbulence, with closer agreement when only the energy contribution from clustered vortices is considered.
\end{abstract}

\maketitle

\section{Introduction}
One of the most fascinating far-from-equilibrium phenomena in classical physics is fluid turbulence. Turbulence can be studied using the incompressible Navier-Stokes equation~\cite{landau1987fluid},
\begin{equation}
\left( \partial_t + \bm v \cdot \bm \nabla \right) \bm{v} = - \bm{\nabla} P + \nu \nabla^2 \bm{v} + \bm{f}_{\textrm{ext}}.
\end{equation}
Here, $\bm{v}$ is the velocity of the fluid, $\nu$ is the kinematic viscosity while $P$ and $\bm{f}_{\textrm{ext}}$ are the pressure and external forcing, normalized by the density of the fluid. Conservation of mass density implies that the fluid velocity is divergence-free in the incompressible limit, $\bm{\nabla}\cdot \bm{v}=0$. The nonlinear advection $(\bm{v}\cdot \bm{\nabla})\, \bm{v}$ is ultimately responsible for the complexity associated with turbulence. The relative importance of nonlinear advection and viscous dissipation is measured by the Reynolds number, $\mathrm{Re}=v_0L/\nu$, with characteristic velocity and length scales $v_0$ and $L$. At small $\mathrm{Re}$, viscosity dominates and the flow is typically laminar. At sufficiently large $\mathrm{Re}$, the flow becomes turbulent~\footnote{The precise value of $\textrm{Re}$ where turbulent fluid motion ensues depends on the shape and boundary conditions of the fluid and is not universal}, with interacting eddies spanning over a range of length scales~\cite{van1982album}.

A defining feature of turbulence is a kinetic-energy \emph{cascade}. A cascade refers to a directed transfer of energy across scales through nonlinear coupling between Fourier modes of the velocity field~\cite{richardson2007weather}. When forcing and dissipation act at widely separated scales, an intermediate \emph{inertial range} can develop in which both play a negligible role. In a statistically stationary state, the mean energy flux within the inertial range is approximately constant across scales~\cite{mccomb1990physics}. Kolmogorov's theory relates this constant flux to a universal scaling of the kinetic energy spectrum~\cite{kolmogorov1991local}. We motivate the kinetic energy spectrum $E(k)$ via~\footnote{A more rigorous derivation of the Kolmogorov scaling law comes from the behavior of the velocity equal time correlator $C(\bm r) = 1/2 \, \langle \bm v(\bm r) \cdot \bm v(0) \rangle$ under the proviso of spatial homogeneity and isotropy, where we define $C(0) \equiv \int \dd k \, E(k)$. In wavenumber space, $E(k)$ can be written more explicitly and is given by $E(k)=(S_{d-1}k^{d-1})\langle |\bm{v}(\bm{k})|^2\rangle/(2 (2\pi)^dV)$ where $S_{d-1}$ is the area of the sphere in $d-$dimensions, $V$ is the spatial volume of the system and $\bm{v}(\bm{k})$ is the Fourier transform of the velocity $\bm{v}(\bm{r})$. }
\begin{equation}\label{eq_kinetic_energy}
 \frac{1}{2V}\left\langle\int \dd^d r\,
 \bm v^2(\bm r,t)\right\rangle
 =\int_0^\infty \dd k\,E(k),
\end{equation}
where $d$ is the spatial dimension, $V$ is the volume of the system and $\langle.\rangle$ denotes a time average. Assuming local homogeneity and isotropy, and that the inertial-range spectrum depends only on the wavenumber $k$ and a scale-independent energy flux $\varepsilon$ (energy transfer rate  per unit mass), dimensional analysis yields
\begin{equation}
 E(k)=C\varepsilon^{2/3}k^{-5/3},
\end{equation}
where $C$ is a dimensionless constant of order unity.

In this article, we are concerned with the turbulent dynamics of superfluids.
Although superfluidity is intrinsically a quantum phenomenon, many aspects of superfluid turbulence closely resemble their classical counterparts, despite the absence of conventional viscous dissipation in a zero temperature superfluid~\cite{volovik2003classical,barenghi2014introduction}. A superfluid is described by a macroscopic complex order parameter $\psi=\sqrt{\rho}\exp(i\theta)$, where $\rho$ is the density and $\theta$ the phase~\cite{anderson1966considerations,madelung1927quantentheorie}. The phase determines the superfluid velocity ($\bm v$) and current ($\bm J$) through ($\hbar=m=1$)
\begin{equation}
 \bm J=\rho\bm v=\rho\bm\nabla\theta,
\end{equation}
where $m$ is the boson mass. Single-valuedness of the order parameter implies quantized circulation, $\oint_{\mathcal C}\bm v\cdot\dd\bm\ell=2\pi q$, with winding number $q\in \mathbb{Z}$.  Unlike classical fluid flow, a superfluid carries vorticity only through quantized vortices, making these topological defects the fundamental degrees of freedom underlying turbulent flow. In a seminal paper, Feynman conjectured that turbulence in 3$d$ superfluids arises from a tangle of quantized vortex lines, whose continual reconnections sustain the turbulent cascade~\cite{feynman1955chapter}. Indeed, turbulent cascades exhibiting approximate Kolmogorov scaling have been observed in superfluid Helium~\cite{maurer1998local} and, more recently, in ultracold atomic gases~\cite{navon2016emergence,zhao2025kolmogorov}.

At sufficiently low temperatures, where the normal component is negligible, the dynamics of a weakly interacting Bose superfluid are described by the Gross-Pitaevskii (GP) equation~\cite{gross1961structure,pitaevskii1961vortex},
\begin{equation}\label{GP_equation}
\textrm{i} \partial_t \psi = - \frac{1}{2}\nabla^2 \psi - \mu \psi + g \vert \psi \vert^2 \psi.
\end{equation}
where $\mu$ denotes the chemical potential, and $g$ the strength of short-range repulsive interactions ($g$ is dimensionless in 2$d$). The GP equation captures both quantized vortices and density fluctuations, providing a framework for studying their coupled dynamics in superfluid turbulence. Although energy can be exchanged between vortex motion and sound, the Hamiltonian equation contains no explicit damping. Dissipation is commonly introduced through the replacement $\partial_t\rightarrow(1+\textrm{i}\gamma)\,\partial_t$, where $\gamma$ is a phenomenological, possibly scale-dependent damping parameter~\cite{choi1998phenomenological,tsubota2002vortex,numasato2010possibility}.

Numerical simulations of the GP equation have reported Kolmogorov scaling in the incompressible kinetic-energy spectrum in both two~\cite{reeves2013inverse,numasato2010direct,shinn2026spontaneous} and three spatial dimensions~\cite{nore1997kolmogorov,kobayashi2005kolmogorov,kobayashi2005kolmogorov-2}. Complementary studies characterize turbulence through circulation statistics~\cite{massaro2025circulation}. Since the kinetic-energy density is $\rho\bm v^2/2$, it is convenient to introduce the density-weighted velocity field $\bm u=\sqrt{\rho}\,\bm v$, whose squared magnitude gives twice the kinetic-energy density. We decompose this field into compressible and incompressible components, $\bm u=\bm u_c+\bm u_i$ so that $\bm\nabla\times\bm u_c=0$ and $\bm\nabla\cdot\bm u_i=0$. In analogy with Eq.~\eqref{eq_kinetic_energy}, we define the incompressible kinetic-energy spectrum $E_i(k)$ by
\begin{equation}
\frac{1}{2} \left\langle \int \dd^d r \ u_i^2(r) \right\rangle \equiv \int_0^\infty \dd k  \, E_i(k).
\end{equation}
Numerical simulations of the GP equation in 2$d$ have identified turbulent regimes with an incompressible kinetic-energy spectrum of the form~\cite{bradley2012energy,reeves2013inverse}
\begin{equation}
E_i(k)\sim
  \begin{cases}
    k^{-5/3}, & k\xi \ll 1,\\
    k^{-3}, & k\xi \gg 1.
  \end{cases}
\end{equation}
Here, the healing length $\xi=1/\sqrt{2g\rho_0}$ characterizes the distance over which the density recovers from zero at the vortex core to its bulk value $\rho_0$. The low-$k$ regime reflects collective vortex dynamics, whereas the high-$k$ tail arises from the vortex-core density profile~\cite{reeves2013inverse} and does not represent an energy cascade (Appendix~\ref{app:dipole_scaling}).

Unlike three-dimensional turbulence, where energy typically flows toward smaller scales, two-dimensional incompressible turbulence supports an inverse energy cascade toward larger scales~\cite{kraichnan1967inertial,kraichnan1980two,paret1997experimental}. In compressible two-dimensional superfluids, both inverse and direct energy cascades have been reported~\cite{reeves2013inverse,numasato2010direct,chesler2013holographic,jiang2026nonequilibrium}.
In connection with the inverse energy cascade, a statistical tendency of like-signed vortices to cluster in 2$d$ was identified by Onsager~\cite{onsager1949statistical}. Experiments in two-dimensional quantum fluids have observed the formation of like-signed vortex clusters and large-scale flow~\cite{neely2013characteristics,gauthier2019giant,johnstone2019evolution,reeves2022turbulent,panico2023onset}. Numerical studies using effective vortex models connect this clustering to inverse energy transfer~\cite{chesler2014vortex,skaugen2017origin,billam2015spectral}. Numerical GP studies further demonstrate spontaneous vortex clustering, with the formation depending on vortex loss, sound emission, and boundary conditions~\cite{simula2014emergence,kanai2021true}

In this work, we introduce a novel approach to superfluid turbulence that takes the collective dynamics of quantized vortices as its starting point. Motivated by the central role of vortices in incompressible turbulent flow, we apply the boson-vortex duality~\cite{lee1991anyon,arovas1997dynamical,tsaloukidis2025nonequilibrium} to describe driven, dissipative superfluid dynamics far from equilibrium. The duality recasts the superfluid in terms of explicit vortex coordinates coupled to an emergent gauge field, which describes the boson density and current. Rather than prescribing vortex interactions through a fixed background, we evolve the vortices and fields together.
Our formulation allows forcing and dissipation to be specified separately for vortices and fluid fields, and provides a natural description of vortex friction as a dissipative current. More generally, our formulation allows the effective vortex Lagrangian to be modified, for example by including vortex inertia or by promoting the vortex degrees of freedom to a separate dynamical field. This opens a route to studying vortex fluctuations and condensation beyond the point-vortex regime, including near the superfluid-Mott insulator transition in a \emph{lattice} superfluid or in the Mott insulator phase. 

In our analysis, we represent vortices as point particles coupled to the dynamical gauge fields and show that the dual equations of motion directly reproduce superfluid hydrodynamics. We numerically evolve the dual equations while driving the system through stochastic vortex pair production and include dissipation through pair annihilation, mutual friction, and damping of compressible motion. At late times, the system approaches a turbulent steady state. In the nearly incompressible regime, we observe clustering of like-signed vortices, quantified through nearest-neighbor sign correlations and the population of the largest clusters. An approximately constant negative incompressible energy flux over a broad wavenumber range indicates an inverse energy cascade, with vortex transport providing the dominant contribution to the flux. The incompressible kinetic-energy spectrum is consistent with Kolmogorov $k^{-5/3}$ scaling, with closer agreement for the clustered-vortex part. 

\begin{figure*}[!t]
    \centering
    \includegraphics[width=\linewidth]{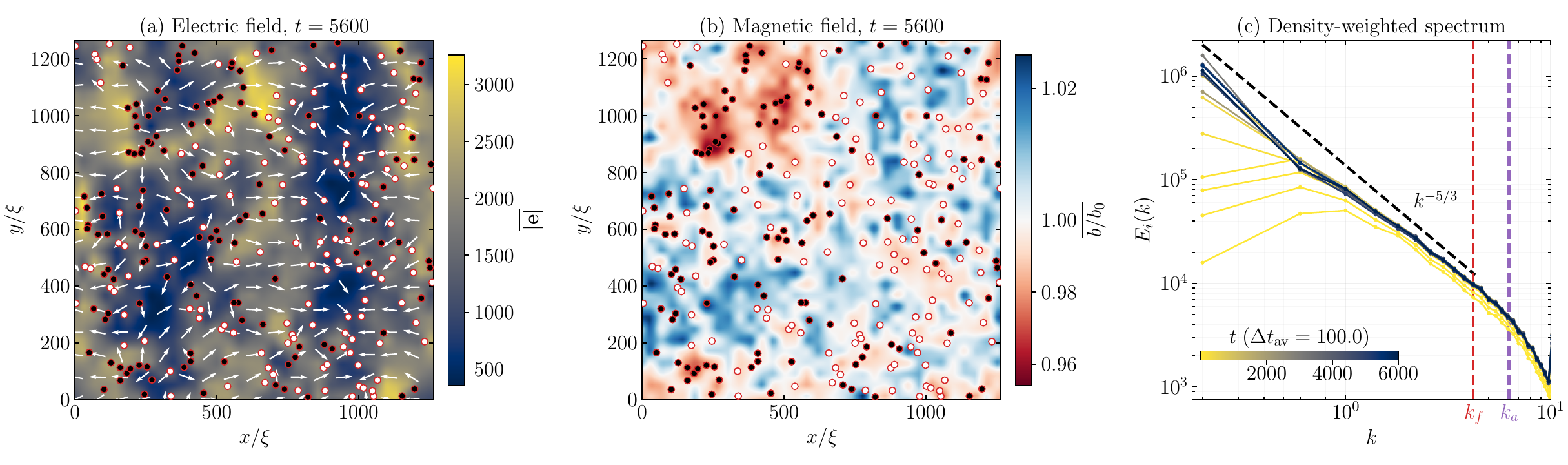}
    \caption{A snapshot of \textbf{(a)} electric fields ($\bm{e}$, or the $\pi/2-$rotated superfluid current)  and \textbf{(b)} normalized magnetic fields ($b/b_0$, where $b_0$ is the average magnetic field, or the 
    superfluid density) at a time $t=1325$ (in units of $\hbar/\mu$), obtained under turbulent dynamics of a two dimensional superfluid with $g \approx 500$ (in units of $\hbar^2/2m$) using the dual hydrodynamical equations (Eqs.~\eqref{eq:Faraday_law}, \eqref{eq:Gauss_law} and \eqref{eq:Ampere_law}). The snapshot is shown at a representative time when the cascade in the incompressible kinetic energy is well developed. 
    For visualization, the color maps in (a,b) are averaged over $4\times4$ grid cells linearly interpolated in both spatial directions between block centers. The point-vortices are depicted by unfilled and filled circles, corresponding to vortices of opposite sign. \textbf{(c)} Incompressible kinetic energy $E_i(k)$, plotted as a function of wavenumber $k$ in a log-log plot at various times (see color bar). At late times, $E_i(k)$ exhibits a power law scaling consistent with Kolmogorov's law $(k^{-5/3})$ for $k< k_f$. $k_f=2\pi/d_f$ is the forcing scale determined by vortex pair production and $k_a=2\pi/d_a$ is the scale at which opposite-signed vortices are being annihilated.} 
    \label{fig:snapshots_and_cascade}
\end{figure*}

\begin{figure*}[!t]
    \centering
    \includegraphics[width=\linewidth]{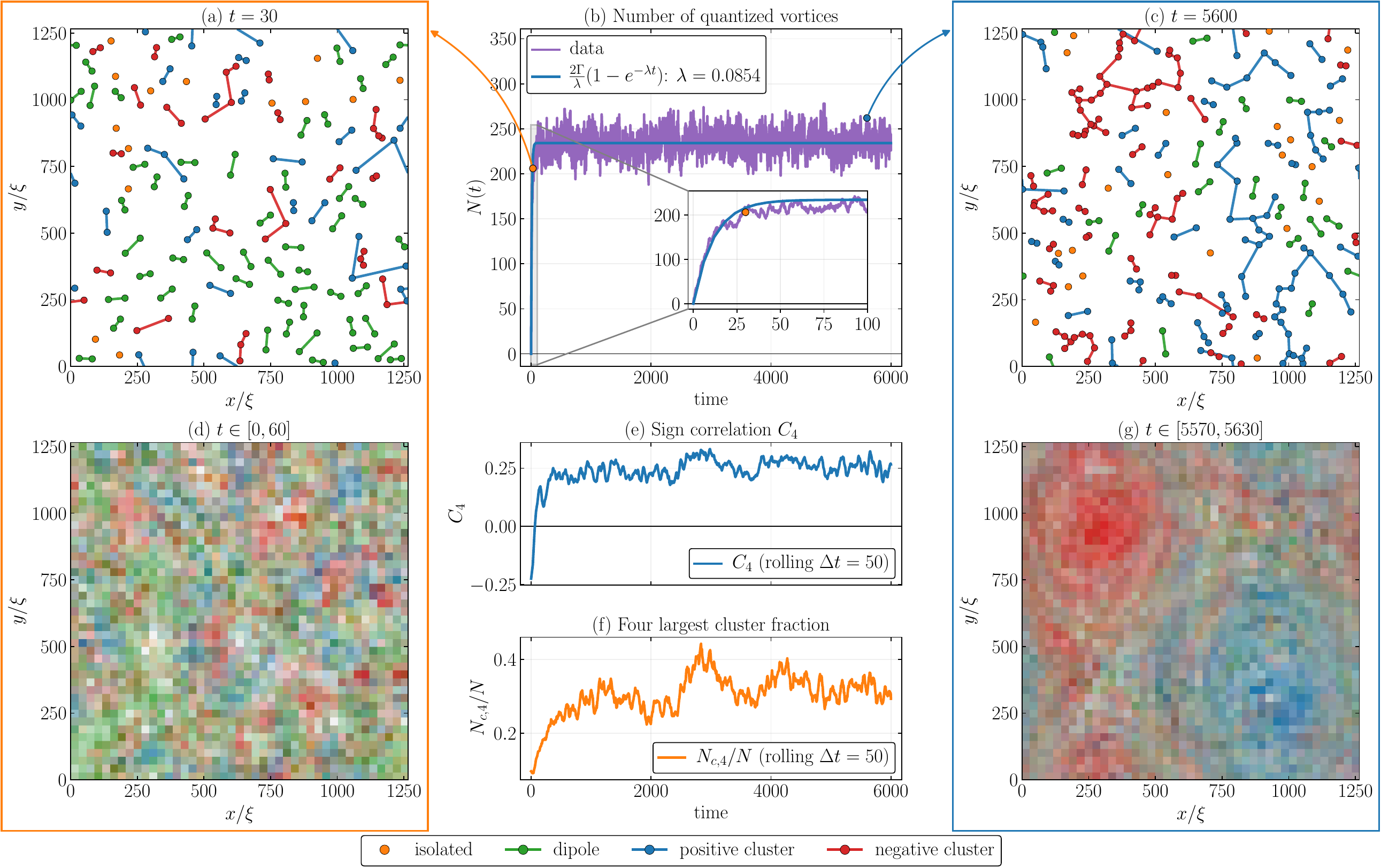}
    \caption{\textbf{(a)} and \textbf{(c)}: Vortex positions at times $t=30$ and $t=5600$, respectively. The vortices have been classified into positive (blue) and negative (red) clusters and dipoles (green), while isolated vortices are colored orange~\cite{reeves2013inverse}. \textbf{(d)} and \textbf{(g)} Corresponding vortex distributions averaged over a time window of 60. \textbf{(b)} Total number of quantized vortices as a function of time, with a fit showing the exponential approach of $N(t)$ to its steady-state value $N_0=2\Gamma/\lambda$. The inset displays the behavior at times $t<100$. \textbf{(e)} Four-nearest-neighbor sign correlation $C_4$ defined in Eq.~\eqref{eq:sign-correlations}. \textbf{(f)} Fraction $N_{c,4}/N$ of the vortices belonging to the four largest same-sign clusters relative to the total number of vortices. In panels \textbf{(e)} and \textbf{(f)}, we take a rolling average over a time window of $\Delta t = 50$. }
    \label{fig:vortices}
\end{figure*}

\section{Hydrodynamics of the dual gauge theory}\label{sec:dual-theory}
A superfluid in $2+1$ dimensional spacetime is characterized by a current 3-vector $J_{\mu} = \left( \rho, \bm J\right)$ [$\mu=0,1,2$], which satisfies the continuity equation $\partial_t \rho + \bm{\nabla} \cdot  \bm J = 0$, or equivalently, $\partial_{\mu} J_{\mu} = 0$.  It follows that $J_{\mu}$ can be written as the curl of a 3-vector $a_{\mu}$, $J_{\mu} = (1/2\pi)\,\epsilon_{\mu \nu \lambda} \partial_{\nu} a_{\lambda}$ ($\epsilon_{\mu\nu\lambda}$ is the Levi-Civita tensor). Clearly, the transformation $a_{\mu} \rightarrow a_{\mu} + \partial_{\mu} \chi$ renders currents invariant, and thus, $a_{\mu}$ acts as a gauge field in $2+1$ spacetime dimensions.  It is the field that mediates the logarithmic interaction between vortices. The density and currents of the superfluid are thus given by,
\begin{eqnarray}
\rho = \frac{1}{2 \pi} (\bm{\nabla} \times \bm{a})_z  \equiv \frac{b}{2 \pi}, \qquad \bm J = \frac{1}{2 \pi} \bm{e}\times \hat{z},
\end{eqnarray}
where $b$ and $\bm e = -\partial_t \bm a + \bm \nabla a_0 $ are the dual magnetic and electric fields respectively. Since our primary interest is the turbulent superfluid velocity, it is also useful to express it directly in terms of the dual gauge fields
\begin{equation}
\bm v(\bm r, t) = \frac{\bm J(\bm r, t)}{\rho(\bm r, t) } = \frac{ \bm{e}(\bm{r},t)\times \hat{z}}{b(\bm r, t)}.
\end{equation}
The incompressible component of the fluid resides in the divergence-free sector of $\bm v$. In the dual theory, the same physics is captured by the curl-free part of $\bm e/b$. In both the bosonic and dual descriptions, the incompressible component of the flow is governed by the vortex degrees of freedom.

The action governing the dual theory involves vortex matter coupled to emergent gauge fields whose dynamics are governed by a \emph{non-relativistic} and \emph{non-linear} version of Maxwell's electrodynamics, 
\begin{align}
\mathcal L_{\textrm{dual}} =  \ &\frac{1}{2 \pi} \left( \frac{1}{2} \frac{\bm e^2}{b} - \frac{g}{4\pi} b^2 -\frac{1}{2}(\grad \sqrt{b})^2 + \mu b  \right) \nonumber \\ &
-[j_v]_{\mu} \,a_{\mu} + \ \hdots
\label{eq:Dual_Lagrangian} 
\end{align}
The minimal coupling $[j_v]_{\mu} a_{\mu}$ involving the vortex 3-current $\displaystyle [j_v]_{\mu}=(\rho_v, \bm{j}^v)$ indicates that vortices carry emergent gauge charge and therefore experience a dual Lorentz force. The remaining terms govern the dynamics of the emergent electromagnetic fields. $\hdots$ denotes a possible kinetic energy term for the vortex matter. The gauge-field dynamics are nonlinear because of the term $\bm e^2/b$. This is because the coefficient of the electric field energy in such a dual Lagrangian, proportional to the inverse superfluid stiffness~\cite{peskin1978mandelstam, dasgupta1981phase,lee1991anyon,fisher1989correspondence}, has been promoted to a \emph{dynamical} field $b(\bm{r},t)$. We show ahead that this enables us to describe superfluid hydrodynamics accurately within the dual formulation. The energy $\propto (\bm{\nabla}\sqrt{b})^2$ yields a cost to density gradients and sets the profile of the density in the vortex core. The term linear in $b$ fixes a uniform background magnetic flux (equivalently, superfluid density), such that $b\gg \delta b$.

\begin{figure}[!t]
    \centering
        \includegraphics[width=\columnwidth]{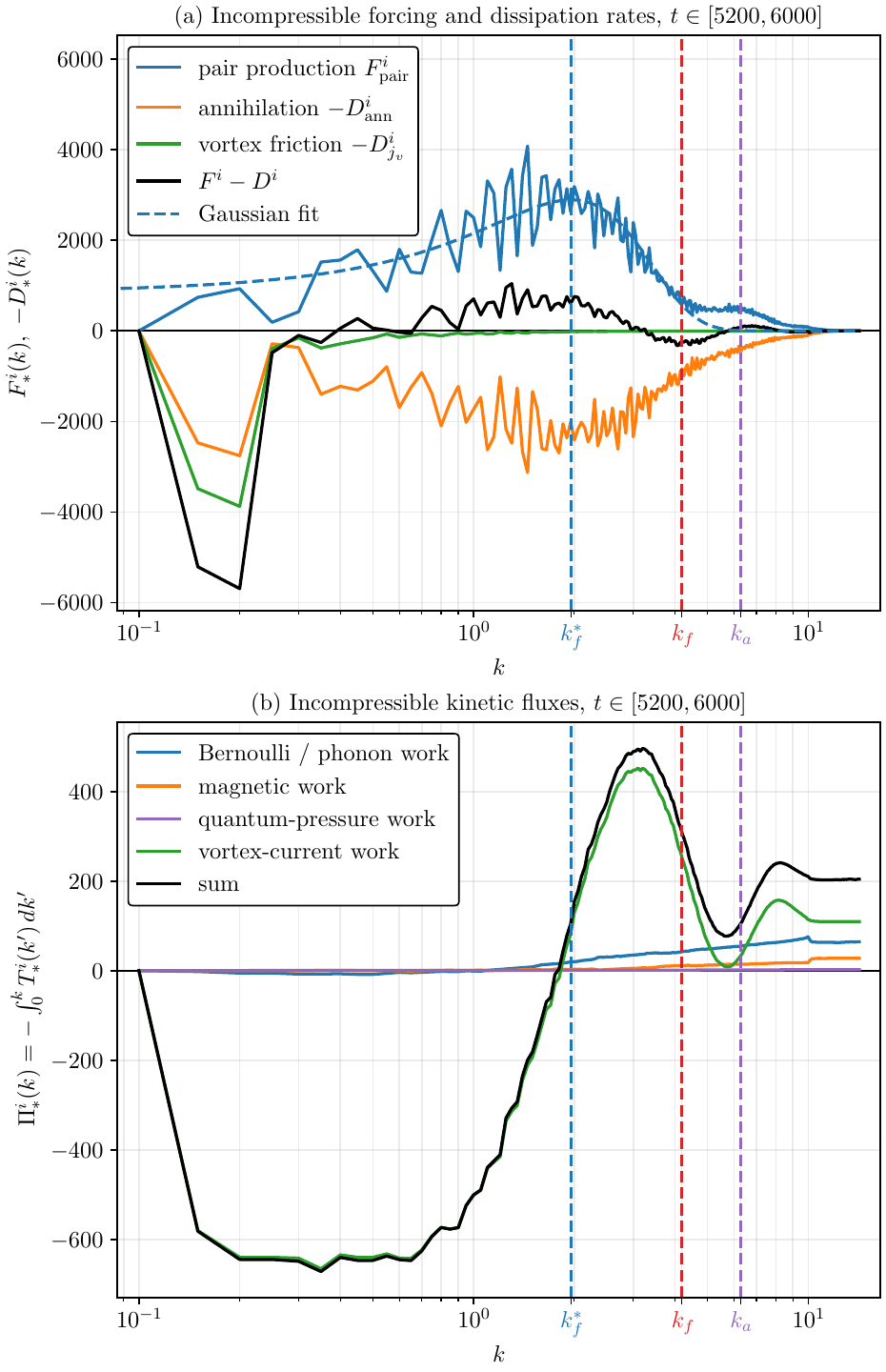}
    \caption{\textbf{(a)} $k$-resolved incompressible forcing and dissipation rates, including vortex pair production $F_\mathrm{pair}^i$, vortex annihilation $D_\mathrm{ann}^i$, and mutual friction $D_{j_v}^i$. The dashed blue Gaussian fit to pair production determines the effective forcing wavenumber $k_f^*$ at its center position. \textbf{(b)} Incompressible conservative energy fluxes, resolved in terms of the individual transfer channels and the total incompressible energy flux. Both panels are averaged over the time window $t\in[5100,5900]$.}
    \label{fig:incompressible_fluxes}
\end{figure}

\noindent The equations of motion of the dual theory are
\begin{subequations}
\begin{alignat}{3}
&\partial_t b \, \hat{z} + \bm{\nabla}\times \bm{e} = 0, \label{eq:Faraday_law} \\ 
&\bm{\nabla}\cdot \left(\frac{\bm{e}}{b}\right) = - 2 \pi \rho_v  ,\label{eq:Gauss_law} \\
&\partial_t \left( \frac{\bm{e}}{b} \right) =\bm{\nabla}\times  \left[\left(\frac{g}{2\pi} \, b + \frac{e^2}{2b^2} + Q(b)
\right)\hat{z}\right] + 2 \pi \bm{j}^v.\label{eq:Ampere_law}
\end{alignat}
\end{subequations}
Eq.~\eqref{eq:Faraday_law} is Faraday's law, which is simply the dual representation of boson density conservation, $\partial_{\mu} J_{\mu} = 0$.  Eq.~\eqref{eq:Gauss_law} is a non-relativistic Gauss law, which encodes the quantization of circulation by identifying point-vortices as sources of the dual electric field. Finally, Eq.~\eqref{eq:Ampere_law} is the non-relativistic version of Ampère's law and governs the dynamics of the dual gauge field with a nonlinear dependence on $\bm e/b$. $Q(b) = -(\nabla^2 \sqrt{b})/(2 \sqrt{b})$, arising from the $(\nabla\sqrt{b})^2$ energy, is a quantum potential that suppresses density gradients~\cite{ballentine1989quantum}. Together, these equations provide a description of the coupled dynamics of vortices and emergent gauge fields. The normalized flow dynamics depend on the background chemical potential $\mu_0=gb_0/(2\pi)$, rather than on $g$ and $b_0$ separately. Although we vary $b_0$ numerically, we report equivalent couplings at the reference density $b_{\rm ref}=2\pi$, for which $g=\mu_0$.

As $g \rightarrow \infty$, dual magnetic field (or equivalently density) fluctuations are effectively frozen out. This realizes incompressibility in the dual theory. In this limit, the dual electric field is longitudinal (Faraday's law) and is completely determined from the instantaneous vortex configuration (Gauss law). As the vortices move, they redistribute electric field energy across length scales. The Kolmogorov cascade can be understood then as a cascade of electric field energies arising from vortex motion. While the electric fields fluctuate intensely, the vortices are weakly coupled, as may be understood by integrating out the gauge fields. This is because, unlike Maxwell's theory, the `coupling constant' $g$ enters the numerator of the coefficient of the magnetic field energy in Eq.~\eqref{eq:Dual_Lagrangian}. 

\begin{figure*}[!t]
    \centering
    \includegraphics[width=\linewidth]{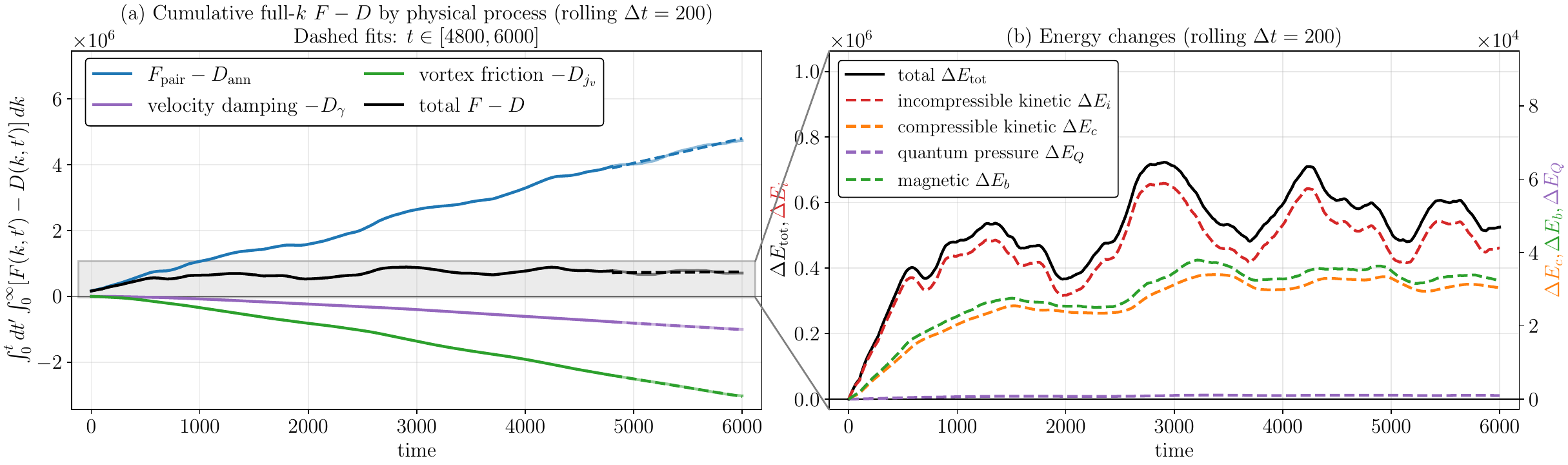}
    \caption{\textbf{(a)} Cumulative forcing and dissipation contributions as a function of time. This includes the difference between vortex pair production $F_\mathrm{pair}$ and annihilation $D_\mathrm{ann}$, mutual friction $D_{j_v}$,  velocity damping $D_\gamma$ and the total energy input $F-D$. Linear fits at late times indicate a constant forcing or dissipation rate. \textbf{(b)} Changes in the different energy components relative to their initial values.}
    \label{fig:energy_steady_state}
\end{figure*}

\section{Numerical simulation with forcing and dissipation}\label{sec:numerics}

We perform numerical simulations of the dual theory. The quantized circulation of the vortices is accounted for by modeling the vortices as point-particles carrying quantized gauge charge. The dynamics of these point vortices are described by a worldline action (Appendix~\ref{app:Dual_theory}), with vortices minimally coupled to the gauge field via $j_\mu^v a_\mu $. The Lagrangian realizing this theory is then Eq.~\eqref{eq:Dual_Lagrangian} without any additional terms. Because the vortex density is typically much smaller than the background magnetic flux density $b_0 = 2\pi \rho_0$, their dynamics is that of guiding centers of charged particles in the lowest Landau level~\cite{ao1993berry}. The vortices then source the violently fluctuating electric fields (boson currents). In particular, in this limit, each vortex drifts with the local superfluid velocity evaluated at its position, $\dot{\bm X}_a = \bm v(\bm X_a)$ for vortex $a$~\cite{thouless1996transverse}. This result naturally follows from the Lorentz force in the dual formulation (Appendix~\ref{app:Dual_theory}), but can also be obtained from the superfluid action in boson coordinates, as shown in Appendix~\ref{app:Guiding_center}. 

We numerically simulate the dual theory by evolving its equations of motion in~\eqref{eq:Faraday_law} and~\eqref{eq:Ampere_law} subject to Gauss' law~\eqref{eq:Gauss_law}. The dual fields are represented on a periodic spatial grid, while vortices move at continuous positions. At each timestep, the field equations and vortex trajectories are solved self-consistently by iterating field and particle updates, with the vortex current deposited along the trial trajectories. Evolution at each timestep is accepted once the field and particle residuals satisfy prescribed convergence tolerances. The vortex current is deposited such that the discrete continuity equation holds to roundoff, thereby preserving Gauss' law up to the nonlinear solver's tolerance. Further details of the numerical algorithm are given in Appendix~\ref{app:numerical_algorithm}.

To induce turbulent superfluid dynamics, we implement stochastic vortex pair production and annihilation. Pair production provides a forcing mechanism, simulating vortex nucleation from obstacles moving through a superfluid. Dipole annihilation happens generically in a superfluid, and is a mechanism for the conversion of vortex energy into phononic excitations. Vortex-antivortex dipoles of unit charge are stochastically injected~\cite{skaugen2017origin} into the fluid at random positions and orientations, with a fixed initial separation $d_f =1.5$ ($ \approx 47.4 \xi$). The number of generated vortex pairs after each timestep $\Delta t$ is drawn from a Poisson distribution, $\Delta N_{\rm pair} \sim \text{Poisson}(\Gamma \Delta t)$ with a nucleation rate $\Gamma = 10$. A vortex and an antivortex annihilate when their separation is below a threshold $d_a = 1.0$ ($\approx 31.6\xi)$ $<d_f$. We associate the forcing and dissipation scales $k_f = 2\pi/d_f$ and $k_a = 2\pi/d_a$ with vortex pair production and annihilation, respectively.

Since $d_a<d_f$, considering isolated dipoles, one can estimate a net energy input proportional to $\ln(d_f/d_a)$ for a combined pair production and annihilation process. Energy therefore continues to enter the system even when the production and annihilation rates balance. To obtain an energy steady state, we hence include further dissipation mechanisms. Compressible motion is damped phenomenologically through
\begin{align}
 \left.\partial_t\bm v \right|_\mathrm{diss} =-\gamma\,\bm v_c,
\end{align}
where $\gamma$ (not to be confused with the previous $\gamma$ damping in the GPE) models relaxation of sound-like excitations into unresolved degrees of freedom. Directly damping the vortex-induced incompressible velocity in the same way would violate Gauss' law. Any forcing or dissipation to the incompressible part must hence be introduced through vortex motion itself. We model thermal dissipation phenomenologically by treating the normal fluid component as a stationary thermal background~\footnote{By assuming a stationary normal component, $\bm v_n = 0$, we select a preferred frame for superfluid dynamics and break Galilean invariance.}, motivated by Landau's two-fluid picture~\cite{landau1941theory} (Appendix~\ref{app:mutual_friction}). We neglect the dynamics of the normal fluid and reactive friction, resulting in the vortex drift law~\cite{hall1956rotation,skaugen2017origin,mehdi2023mutual,kobayashi2006thermal}
\begin{align}
\label{eq:vortex_drift_law} 
\dot{\bm X}_a=\bm v(\bm X_a)-\alpha q_a\hat{\bm z}\times\bm v(\bm X_a),
\end{align}
where $\alpha$ characterizes the dissipative coupling to the thermal background. The drift law~\eqref{eq:vortex_drift_law} motivates the decomposition of the vortex current in terms of a conservative and a dissipative part, $\bm j_v = \bm j_v^\text{adv} + \bm j_v^\text{diss}$, with $j_v^\text{adv} = \rho_v \bm v$. In the local continuum description, we interpret the dissipation mechanism as an Ohm's law-like contribution to the dual vortex current in Eq.~\eqref{eq:Ampere_law},
\begin{equation}
 \bm j_v^{\mathrm{diss}}=-\sigma_v\bm e,\qquad
 \sigma_v=\frac{\alpha n_v}{b}\geq0,
 \label{eq:main-vortex-conductivity}
\end{equation}
where $n_v=\sum_a\delta(\bm x-\bm X_a)$ is the unsigned vortex density.
Vortex motion hence mediates relaxation of the physical boson current $\bm J=\rho\bm v$, with the dissipative contribution reading (Appendix~\ref{app:mutual_friction}) 
\begin{equation}
 \left.\partial_t\bm J\right|_{\mathrm{diss}}
 =-2\pi\alpha n_v\bm J.
 \label{eq:main-mutual-friction-current-relaxation}
\end{equation}
The linear damping ($\gamma$) suppresses acoustic motion, while mutual friction ($\alpha$) dissipates kinetic energy through vortex transport and reduces the separation of isolated vortex-antivortex pairs, promoting their eventual annihilation. Neither term contains the explicit high-wavenumber weighting of viscous dissipation ($\propto \nabla^2\bm{v}$). Instead, their spectral effects depend on the evolving flow and vortex distribution. We obtain a turbulent energy cascade for weak dissipation, choosing $\alpha=5\times10^{-4}$ and $\gamma=3\times10^{-3}$ in our numerics, but our results do not sensitively depend on their precise values.

\section{Vortex clustering}\label{sec:vortices}

The dual equations are evolved on a torus of size $40\times 40$ ($\approx 1265\xi \times 1265 \xi$) on a grid containing $128\times 128$ points. To minimize the effects of compressibility, we choose $g=500$. After $\sim10^3$ time units, we find that the superfluid undergoes turbulent dynamics. In \cref{fig:snapshots_and_cascade}(a) and (b), we provide representative snapshots of the electric and magnetic field configurations respectively, of when the fluid is undergoing a turbulent cascade. In \cref{fig:snapshots_and_cascade}(c), we plot the incompressible kinetic energy spectrum $E_i(k)$ as a function of wavenumber as it evolves in time. At late times, we find a distinct scaling regime emerge, with a power law that is approximately consistent with an exponent of $-5/3$. The scaling regime is most pronounced at wavenumbers less than $k_f$. 

To further characterize the turbulent state, we analyze the vortex behavior. In \cref{fig:vortices}(b), we plot the number of vortices as a function of time. The number of vortices increases in the simulation over time, until production and annihilation rates balance each other. We model the vortex number as a function of time through the differential equation 
\begin{align}
    \label{eq:vortex_diff-eq}
    \partial_t N(t) = 2\Gamma - \lambda N(t).
\end{align}
Numerical evidence for the annihilation rate being proportional to $N(t)$ is given in Appendix~\ref{app:vortex_steady_state}. Eq.~\eqref{eq:vortex_diff-eq} is solved by $N(t) = 2\Gamma/\lambda \, (1-e^{-\lambda t})$. A fit to the data in \cref{fig:vortices}(b) determines $\lambda = 0.08540$. With $\Gamma = 10$, this implies a steady state vortex number $N_0 = 2\Gamma/\lambda = 234.19$, which translates to an unsigned average vortex density of $\bar{n}_v \simeq 1.46\times 10^{-4} \xi^{-2}$. The characteristic timescale $\tau = 1/\lambda \simeq 11.71$ of the exponential relaxation is much smaller than the time scale associated with the onset of turbulent dynamics. In \cref{fig:vortices}(a) and (c), we provide snapshots of the vortex positions at early and late times of our dynamics, classified in terms of positive and negative clusters, dipoles and isolated vortices~\cite{reeves2013inverse} (Appendix~\ref{app:clustering}).

Turbulent dynamics of a 2$d$ superfluid is believed to induce vortex \emph{clustering}~\cite{reeves2013inverse,billam2014onsager,skaugen2016vortex,skaugen2017origin}. We find vortex distributions consistent with this expectation (see Appendix~\ref{app:clustering} for details on how the vortex distribution is obtained). In particular, we find the emergence of macroscopic clusters at late times during the turbulent dynamics. This is particularly evident in \cref{fig:vortices}(g), which shows the vortex distribution averaged over a time window of 60 units. We identify two major regions of clustering in \cref{fig:vortices}(g), one with positive and one with negative charge. This is especially striking when comparing to \cref{fig:vortices}(d), where the vortex distribution at an early time is seen to be dominated by dipoles. To quantify the amount of vortex clustering, we compute the average sign correlation~\cite{white2012creation,reeves2013inverse,yang2025quantum} (further details in Appendix~\ref{app:clustering})
\begin{equation}
 C_p=\frac{1}{p N}\sum_{a=1}^{N}\sum_{j=1}^{p}q_aq_{a_j}
 \label{eq:sign-correlations}
\end{equation}
of the four nearest neighbors, $p=4$, shown in \cref{fig:vortices}(e). Positive correlations indicate a preference for like-signed neighbors, while negative correlations indicate
opposite-sign pairing. While a random vortex distribution implies $C_4 = -1/(N-1)$, we observe an initial negative sign correlation characterizing the dipole regime, which increases and becomes positive close to the time $t_0$ where the steady state in vortex number is reached. It further increases to a value of $C_4 \approx 0.25$, which implies that each vortex on average has one more same-signed than opposite-signed vortex among its four nearest neighbors. This is further strengthened by \cref{fig:vortices}(f), which shows that the fraction of vortices $N_{c,4}/N$, participating in the four largest clusters, increases as a function of time. Together, these measures indicate that vortices form clusters containing an increasing number of vortices over time, consistent with energy transfer from small to large scales.

\section{Energy fluxes and turbulent steady state}\label{sec:steady-state}

It is likely that vortex clustering is accompanied by suppressed dipole annihilation. This suggests an emergent dynamical regime of conservation of vortex number (even though the dynamics is manifestly non-conserving), which in turn, through Gauss' law in Eq.~\eqref{eq:Gauss_law}, implies a conservation of vorticity $(\bm{\nabla}\times \bm{v})_z$. In 2$d$, it is well known that the conservation of the square of vorticity---\emph{enstrophy}---in an incompressible fluid causes an \emph{inverse} cascade of energy towards smaller wavenumbers~\cite{fjortoft1953changes,kraichnan1967inertial, kraichnan1980two}. Even though the superfluid has a finite compressible component, vortex clustering in our dynamics suggests the existence of an inverse cascade. 

\begin{figure}[!t]
    \centering
    \includegraphics[width=\columnwidth]{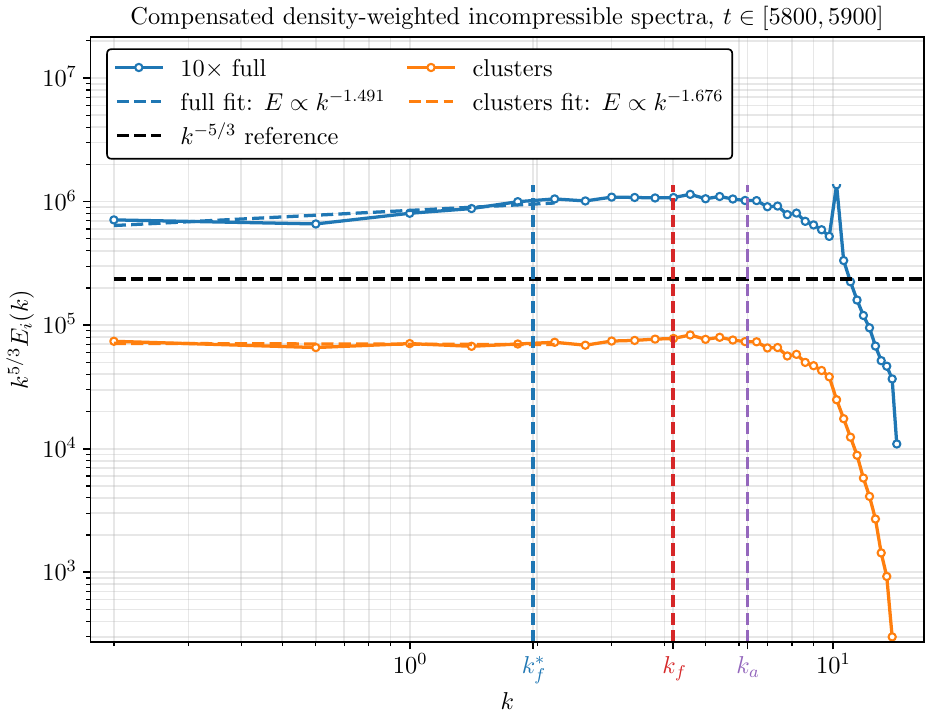}
    \caption{Compensated incompressible kinetic energy spectrum at late times corresponding to the clustered state, offset by a factor of $10$ for visualization and shown in blue. In orange, we show the incompressible spectrum obtained from the clustered vortices only, weighted by the full magnetic field. Both curves are multiplied by $k^{5/3}$, such that the Kolmogorov scaling manifests as a horizontal line. Linear fits to the data for $k\leq k_f^*$ are visualized by dashed lines.}
    \label{fig:cluster_spectrum}
\end{figure}

To quantitatively determine the flow of energy, we decompose the change of the incompressible energy spectrum (label $i$) at wavenumber $k$ as
\begin{align}
    \partial_t E_i(k,t) = T^i(k,t) + F^i(k,t) - D^i(k,t).
\end{align}
Here, $F^i(k,t)$ and $D^i(k,t)$ denote external forcing and dissipation that may inject or drain energy from the system, while $T^i(k,t)$ represents a conservative energy transfer that merely redistributes energy between different modes.
This redistribution can involve compressible and incompressible kinetic energy, magnetic-field (interaction) energy, and quantum-pressure energy. \Cref{fig:incompressible_fluxes}(a) shows the forcing and dissipation spectra in the incompressible sector: pair production $F_\mathrm{pair}^i$, dipole annihilation $-D_\mathrm{ann}^i$, and mutual friction $-D_{j_v}^i$. The pair-production and annihilation rates are obtained from changes in the incompressible energy spectrum across vortex insertion and removal steps (Appendix~\ref{app:fluxes}). Both contributions extend over broad wavenumber bands rather than being localized at a single wavenumber. A Gaussian fit to the pair-production spectrum defines an effective forcing wavenumber $k_f^*<k_f$. This spectral injection scale is measured from the evolving system rather than prescribed directly by the injection protocol. Mutual-friction losses are concentrated at small $k$ with a spectral distribution that dynamically emerges from the coupled field-vortex dynamics without an explicitly imposed large-scale damping filter. Such a distribution is consistent with energy transfer toward large scales, where mutual friction provides an energy sink.

In order to diagnose the presence of an energy cascade, we consider the conservative incompressible flux obtained from the transfer rate $T^i(k,t)$ and decompose it in terms of four distinct flux channels (Appendix~\ref{app:fluxes}). From the conservative equation of motion of the density weighted velocity $\bm u$, 
\begin{align}
 \partial_t\bm u
 ={}&-\sqrt{\rho}\,\nabla
       \left(\frac{|\bm u|^2}{2\rho}\right)
     -\frac{\bm u}{2\rho}\,
       \nabla\cdot(\sqrt{\rho}\bm u)
 \nonumber\\
 & -g\sqrt{\rho}\,\nabla\rho-\sqrt{\rho}\,\nabla Q(\rho) -2\pi\sqrt{\rho} \, \hat{z} \times \bm j_v^\text{adv}
\end{align}
we obtain the `Bernoulli / phonon (p)' channel from the terms in the first line, the 'magnetic (g)' and 'quantum-pressure (Q)' channels from $-g\sqrt{\rho}\,\nabla\rho$ and $-\sqrt{\rho}\,\nabla Q(\rho)$, respectively as well as the 'vortex-current (v)' channel from $-2\pi\sqrt{\rho} \,\hat{z} \times \bm j_v^\text{adv}$. For each channel, we apply the incompressible Helmholtz projector to the corresponding terms and take their spectral inner product with $\bm u_i$ to obtain the transfer rate $T_\chi^i(k,t)$. Each channel satisfies the energy-conservation sum rule $\sum_a\int_0^\infty T_\chi^a(k,t)\,\mathrm dk=0$, where $a$ runs over the participating energy sectors.
In \cref{fig:incompressible_fluxes}(b), we plot the fluxes 
\begin{align}
    \Pi_\chi^i(k) = \left\langle -\int_0^k \dd k^\prime \,T_\chi^i (k^\prime,t) \right\rangle,
\end{align}
with $\chi=\{$p, g, Q, v$\}$ denoting the flux channels. $\langle.\rangle$ denotes an average over a time window of $\Delta t$, where $\Delta t = 800$ here. Within the scaling regime for $k<k_f^*$, the flux is negative and approximately constant. This observation is consistent with Kolmogorov's theory predicting a scale-independent energy flux $\epsilon$, here realized by $\Pi^i(k) = \sum_\chi \Pi^i_\chi(k)$. Our results suggest the existence of an inverse cascade of incompressible kinetic energy in the turbulent dynamics of the 2$d$ superfluid.

In \cref{fig:incompressible_fluxes}(b), the dominant contribution to $\Pi^i(k)$ originates from the vortex-current channel $\Pi^i_\mathrm{v}(k)$. In the continuum description, $\bm j_v^\mathrm{adv}=\rho_v\bm v=\omega\bm v/(2\pi)$, where $\omega=(\nabla\times\bm v)_z$ (c.f. Gauss' law~\eqref{eq:Gauss_law}). Vortex motion thus provides a nonlinear coupling between vorticity and velocity that contributes to the observed energy redistribution across scales.

For wavenumbers $k>k_f^*$, we observe a positive incompressible flux, suggestive of a direct cascade in this regime. Since vortices are injected without the corresponding core-density depletion, the subsequent density relaxation may excite short-wavelength compressible modes and accordingly generate a forward energy transfer.

In \cref{fig:energy_steady_state}(a), we show the cumulative energy input and losses from the different forcing and dissipation mechanisms. Once pair production and annihilation balance statistically, their combined energy injection rate $F_\mathrm{pair} - D_\mathrm{ann}$ is approximately constant (illustrated by the linear fit in blue), with the isolated-dipole estimate predicting a rate proportional to $\Gamma\ln(d_f/d_a)$. While the mean vortex number remains stationary, continued injection increases the incompressible kinetic energy per vortex. This energy accumulation can increase the velocities experienced by the vortices, thereby enhancing mutual-friction losses $(-D_{j_v})$ until they approximately balance the net energy input. Energy reaches a steady state on a longer timescale than vortex-number equilibration and the onset of turbulence, as indicated by the approximately zero slope of the black fit to the cumulative net energy input $F-D$. \Cref{fig:energy_steady_state}(b) shows the resulting changes in the individual energy contributions. Incompressible kinetic energy dominates these changes, while the compressible contribution remains small at the chosen value of $g$.

In \cref{fig:cluster_spectrum}, we compare the incompressible energy spectrum $E_i(k)$ to the spectrum $E_i^\mathrm{clusters}(k)$ reconstructed using only vortices classified as belonging to clusters, excluding dipoles and isolated vortices. Linear fits on the log-log scale over the same wavenumber interval yield an exponent closer to the Kolmogorov prediction of $-5/3$ for the clustered-vortex spectrum. This suggests that vortex clustering is associated with Kolmogorov scaling, while unclustered vortices and their correlations with clustered vortices contribute to deviations in the full spectrum~\cite{skaugen2017origin,billam2015spectral}.

\section{Hydrodynamics in boson coordinates}\label{sec:hydrodynamics}

The emergence of hydrodynamics in the dual theory is not mere coincidence. The underlying reason is that both the GP equation and the dual theory reduce to the same hydrodynamic description. The GP equation can be recast in hydrodynamic form by again writing $\psi = \sqrt{\rho} e^{\textrm{i} \theta}$, and decomposing the phase gradient into smooth and vortex parts, $\bm \nabla \theta = \bm \nabla \theta_\mathrm{sm} +\bm \nabla \theta_v$, where $\bm{\nabla}\times \bm{\nabla}\theta_\textrm{sm}=0$ and $\bm{\nabla}\cdot \bm{\nabla}\theta_v=0$. The imaginary part reduces to the bosonic continuity equation, $\partial_\mu J_\mu =0$, while the real part yields the inviscid Navier-Stokes equation (Appendix~\ref{app:Hydrodynamics_from_GPE}),
\begin{equation}\label{eq:Hydrodynamic_GPE}
\left( \partial_t + \bm v \cdot \bm{\nabla} \right) \bm v = -\bm{\nabla} \left[ Q(\rho) + g \rho \right] + \bm f_\mathrm{Lorentz}.
\end{equation}
The Lorentz force term  $\bm f_\mathrm{Lorentz} = 2\pi \, \hat{z} \times (\rho_v \bm v - \bm j_v)$ is sourced only by the vortices~\footnote{For point-vortices in the GP equation, the Lorentz force vanishes. This is because the vortices are in the lowest Landau level limit and hence $\bm{j}_v=\rho_v\bm{v}$.} The vortex density and current can be interpreted as background magnetic and electric fluxes experienced by the bosons, as in the Josephson effect. Since this force is localized at the vortex cores, where the superfluid density vanishes, the force density satisfies $\rho \bm f_\mathrm{Lorentz}= 0$. To further show that the dual equations captures the hydrodynamics faithfully, we show that it reproduces the $k^{-3}$ scaling for $k\xi\gg1$ by computing $E_i(k)$ for a single dipole (Appendix~\ref{app:dipole_scaling}). Our numerical grid size is unable to resolve the $k^{-3}$ regime at large $g$.

The dual formulation reproduces the same hydrodynamics \emph{more directly}. Unlike the GP equation, no additional change of variables is required: the hydrodynamic equations are already encoded in the equations of motion. In particular, the dual Ampère's law becomes the inviscid Navier-Stokes equation upon substituting $\bm v = - \hat z \times \bm e/b$ (Appendix~\ref{app:Dual_theory}). From the dual perspective, the nonlinear advection in the Euler equation is a direct consequence of the nonlinear gauge-field dynamics encoded in the non-relativistic gauge sector of the dual Lagrangian. 

\section{Discussion}\label{sec:discussion}
We have shown that the dual formulation faithfully reproduces the characteristic features of two dimensional superfluid turbulence, including Kolmogorov scaling.  We have also shown explicitly that the dual equations of motion are equivalent to the hydrodynamic equations obtained from the Gross-Pitaevskii theory. The dual formulation therefore provides a framework in which turbulent cascades can be analyzed directly in terms of vortex degrees of freedom, while also accounting for the effects of compressiblity.

Our calculations using the dual gauge theory have uncovered several new directions of exploration. The turbulent dynamics obtained can be further characterized---such as a comparison of decaying and steady-state turbulence and an analysis of the vortex distributions in the turbulent state~\cite{bradley2012energy}. Given that both direct~\cite{numasato2010direct} and inverse cascades~\cite{reeves2013inverse} have been observed in the 2$d$ compressible superfluids, one can use the ability to manipulate the vortices directly in the dual theory to re-examine this question more systematically. In particular, one may obtain dynamical crossovers or transitions between the two cascades by tuning the compressibility~\cite{fouxon2023compressible, falkovich2017vortices}. These issues will be addressed in future work.

The dual formulation also involves non-linear equations that appear to be as cumbersome as the GP equation.  One may therefore legitimately ask whether any new physics can be gleaned from it that was not already known from studies of the GP equation.  While we have focused our attention here to turbulent behavior deep within the superfluid phase, perhaps the real advantage of the dual formulation occurs away from this regime~\cite{bhattacharjee2026quantum}.  To see why, it is worth considering a {\it lattice} superfluid in two spatial dimensions.  While the lattice breaks translation symmetry, in a coarse-grained sense, the fluid can still exhibit emergent momentum conservation at length scales large compared to the lattice spacing.  The lattice allows for a quantum phase transition between a superfluid and Mott insulator ground states, which cannot be realized in the continuum GP equation limit of the superfluid.  If the superfluid-insulator quantum phase transition is continuous, there is a divergent length scale below which a continuum description again becomes valid. The dual representation of hydrodynamics in this regime would be in terms of a plasma of vortex antivortex pairs, and perhaps is a more natural set of coordinates to describe far-from-equilibrium behavior near the transition.  

The Mott insulating phase occurs  at stronger interaction strengths in the lattice superfluid.  It has no low energy hydrodynamic modes and can be thought of as a dual vortex condensate.  As a consequence of the Anderson-Higgs mechanism, there are no low energy excitations.  Nevertheless, one can imagine driving the Mott insulator far from equilibrium with a drive that exceeds the lowest energy needed to access collective excitations. In this regime, far-from-equilibrium dynamics can be formulated in terms of a continuum model of a Higgs condensate of vortices with gapped emergent gauge fields.  The extent to which turbulent behavior can be realized in this regime shall be the topic of a forthcoming publication. 

\begin{acknowledgements}
We acknowledge useful conversations with A.~Balatsky, M. P. A. Fisher, S.~K.~Jha, Y.~Li, P.~K.~Mishra, R.~Moessner, N.~O'Dea, V.~Oganesyan, V.~Rosenhaus, M. Tsubota, S.~Sondhi and M. K. Verma. We also thank the participants of the workshop on \emph{Emergent Gauge Theories: Bridging Quantum Matter, Quantum Information, and Fundamental Interactions} at MPI-PKS, Dresden, Germany for inspiring discussions. TH, SB and SR are supported in part by the US Department of Energy, Office of Basic Energy Sciences, Division of Materials Sciences and Engineering, under Contract No.~DE-AC02-76SF00515S. TH was also supported by the Deutsche Forschungsgemeinschaft (DFG, German Research Foundation) under Project No. 537357978.
\end{acknowledgements}

\appendix
\makeatletter
\def\@hangfrom@appendix#1#2#3{%
  #1%
  \@if@empty{#2}{%
    \MakeTextUppercase{#3}%
  }{%
    \MakeTextUppercase{#2}%
    \@if@empty{#3}{}{:\ \MakeTextUppercase{#3}}%
  }%
}
\let\@hangfrom@section\@hangfrom@appendix
\makeatother

\section{Derivation of the dual theory}\label{app:Dual_theory}
We briefly discuss the derivation of the dual theory.  The derivation closely follows that described in Ref.~\cite{lee1991anyon}. We start with a Lagrangian for a 2$d$ superfluid,
\begin{equation}
\label{eq:SF_Lagrangian}
\mathcal L_{\textrm{sf}} = \textrm{i} \psi^* \partial_t \psi - \frac{1}{2}\vert \nabla \psi \vert^2 + \mu \vert \psi \vert^2 - \frac{g}{2} \vert \psi \vert^4+\cdots
\end{equation}
with complex order parameter $\psi$ (setting $\hbar=m=1$), where $\cdots$ denote higher powers of $\psi$ and derivatives which are subleading deep in the superfluid phase.  Writing $\psi = \sqrt{\rho}\, e^{\textrm{i} \theta}$, 
\begin{equation}\label{eq:lsf_density_phase_variables}
\mathcal L = - \rho \partial_t \theta - \frac{\rho}{2} \left( \bm{\nabla} \theta \right)^2-\frac{1}{2}(\bm{\nabla}\sqrt{\rho})^2 + \mu \rho - \frac{g}{2} \rho^2 + \cdots
\end{equation}
where we have dropped a total derivative of $\rho$. We decouple the second term via a Hubbard-Stratonovich identity, introducing a field $\bm J$ corresponding to the spatial currents: 
\begin{equation}
\mathcal L = - \rho \partial_t \theta - \bm J \cdot \bm{\nabla} \theta + \frac{1}{2 \rho} \bm J^2 -\frac{1}{2}(\bm{\nabla}\sqrt{\rho})^2+  \mu \rho - \frac{g}{2} \rho^2 + \cdots \nonumber
\end{equation}
We break up the phase gradients into  smooth and vortex parts, $\partial_{\mu} \theta = \partial_{\mu} \theta_{\textrm{sm}} + \partial_{\mu} \theta_v$ and integrate out the smooth fluctuations.  This produces a constraint $\partial_t \rho + \nabla \cdot \bm J = 0$, which is just the global $U(1)$ current conservation:
\begin{equation}
\mathcal L = - \rho \partial_t \theta_v - \bm J \cdot \bm{\nabla} \theta_v  + \frac{1}{2 \rho} \bm J^2 -\frac{1}{2}(\bm{\nabla}\sqrt{\rho})^2+  \mu \rho - \frac{g}{2} \rho^2 + \cdots \nonumber
\end{equation}
Defining $J_{\mu} = \left( \rho ,\bm J \right)$, the constraint of current conservation is $\partial_{\mu} J_{\mu} = 0$ which can be written in terms of a spacetime curl of unconstrained vector fields $a_{\mu}$: 
\begin{equation}
J_{\mu} = \frac{1}{2 \pi} \epsilon_{\mu \nu \lambda} \partial_{\nu} a_{\lambda}
\end{equation}
where $\epsilon_{\mu \nu \rho}$ is the Levi-Civita antisymmetric tensor.  Note that $J_{\mu}$ is invariant under a ``gauge transformation" $a_{\mu} \rightarrow a_{\mu} + \partial_{\mu} \chi$ for an arbitrary scalar $\chi$.  Thus, $a_{\mu}$ behaves as a gauge field.  Written component-wise, the constraint is
\begin{equation}
J_0 = \rho = \frac{b}{2 \pi}, \ \ \bm J = \frac{\bm{e}\times \hat z }{2 \pi}
\end{equation}
where $b = \partial_1 a_2 - \partial_2 a_1$ is the magnetic field pseudoscalar and $ e_i  = -\partial_t a_i+\partial_i a_t $ are electric fields living in the plane.  After integrating the first two terms by parts, and defining vortex currents $[j_v]_{\mu}$ as
\begin{equation}\label{eq:Def_vortex_current}
[j_v]_{\mu} = \frac{1}{ 2\pi} \epsilon_{\mu \nu \lambda} \partial_{\nu} \partial_{\lambda} \theta_v,
\end{equation}
 we arrive at the dual Lagrangian
\begin{equation}
\mathcal L_{\textrm{dual}} = -[j_v]_{\mu} a_{\mu}  + \frac{1}{ 2\pi} \left( \frac{\bm e^2}{2b} - \frac{g}{4\pi} b^2 -\frac{1}{2}(\bm{\nabla}\sqrt{b})^2+\mu b \right) + \cdots \nonumber
\end{equation}
This is Eq.~\eqref{eq:Dual_Lagrangian} of the main text.

\subsection{Equations of motion}
The equations of motion of the dual theory are readily obtained by varying the action.  Following the standard approach~\cite{landau_volume2}, we do so in two steps.  First, we vary the vortex trajectories while holding the gauge fields fixed, and second, we vary the gauge fields while holding the vortex trajectories fixed.  The former yields the Lorentz force law for vortices and the latter yield a non-relativistic version of Maxwell equations for the gauge fields.  

Start with the vortex degrees of freedom.  The first term in $\mathcal L_{\textrm{dual}}$ is manifestly Lorentz invariant and can be taken to ``live" on a vortex worldline.  We first obtain the equations of motion on the worldline and then take their non-relativistic limit.  Parametrizing the  spacetime interval on the worldline via the scalar $\textrm{d}s$, the vortex part of the action is
\begin{equation}
S_v =-\int \dd^2x \dd t \ [j_v]_{\mu} a_{\mu} = -\int \dd s \left[  a_{\mu} \frac{\dd x^{\mu}}{\dd s} \right],
\end{equation}
where $\dd s = \sqrt{\dd x_{\mu} \dd x_{\mu} }$. Ordinarily, an additional kinetic term for vortices on their worldlines of the form $S_0 = -m_vc \int ds$ would be present; however, such a term is absent in the derivation of the dual action above.  It is effectively as though the vortices have no bare kinetic energy. Upon a variation $x_{\mu} \rightarrow x_{\mu} + \delta x_{\mu}$, the vortex part of the action changes to 
\begin{equation}
\delta S_v = \int \delta x^{\mu} \left[ - f_{\mu \nu} \frac{\dd x^{\nu}}{\dd s}  \right] \dd s 
\end{equation}
and the equations of motion therefore are
\begin{equation}
 f_{\mu \nu} \frac{\dd x^{\nu}}{\dd s} = 0.
\end{equation}
Using $\dd s = \dd t \sqrt{1-\dot{\bm x}^2}$ and taking the non-relativistic limit $v \ll 1$, it follows that the vortices obey the Lorentz force law
\begin{equation}\label{eq:Lorentz_force}
0 = \bm e +  \dot{\bm x} \times \hat z \, b.
\end{equation}
This implies that the vortices are drifted along with the superfluid velocity (as reported in the main text),
\begin{align}\label{eq:vortex-velocity}
    \dot{\bm x} = \frac{\bm{e}\times \hat{z} }{b} = \bm v.
\end{align}
Next, we hold the vortex currents fixed and vary with respect to the gauge fields.  Upon such a variation $a_{\mu} \rightarrow a_{\mu} + \delta a_{\mu}$, field strengths vary as $f_{\mu \nu} \rightarrow f_{\mu \nu} + \delta f_{\mu \nu}$ where $\delta f_{\mu \nu} = \partial_{\mu} \delta a_{\nu} - \partial_{\nu} \delta a_{\mu}$.  Equivalently, $\bm e \rightarrow \bm e + \delta \bm e$ and $b \rightarrow b + \delta b$.  Upon such a variation, we obtain the equations of motion,
\begin{align}
&\partial_i \left( \frac{e_i}{b} \right) = - 2 \pi \rho_v,\nonumber \\
&\epsilon_{ij} \partial_j\left( \frac{e^2}{2 b^2} + g \frac{b}{2 \pi} + Q(b) \right) = -2 \pi [j_v]_i + \partial_t \left( \frac{e_i}{b} \right).
\end{align}
In addition, the Faraday law is automatically encoded as a ``Bianchi identity" $ \epsilon_{\mu \nu \rho} \partial_{\mu} \partial_{\nu} a_{\rho} = 0$,
\begin{equation}
\epsilon_{i j } \partial_i e_j + \partial_t b = 0. 
\end{equation}

\section{Derivation of hydrodynamic equations from the GP equation}\label{app:Hydrodynamics_from_GPE}
We discuss how a hydrodynamic equation can be obtained from the GP equation. This is well known in prior literature. Here, we review this derivation and furthermore undertake a more careful treatment of the vortices to obtain the terms corresponding to the vortex charge and current, encapsulated in $\bm{f}_{\textrm{Lorentz}}= 2 \pi \, \hat{z} \times (\rho_v \bm v - \bm j_v)$, that are present in the hydrodynamic form of the GPE in Eq.~\eqref{eq:Hydrodynamic_GPE} of the main text.

We start from the GP equation ($\hbar=m=1$) 
\begin{align}
    \textrm{i} \partial_t \psi = - \frac{1}{2}\nabla^2\psi - \mu \psi + g \vert \psi \vert^2 \psi
\end{align}
and write $\psi(\bm r, t) = \sqrt{\rho(\bm r, t)} \exp{\textrm{i} \theta(\bm r, t)}$. Taking the imaginary part yields
\begin{align}\label{eq:ImPart_GPE}
    \partial_t \rho + \bm{\nabla} \cdot (\rho \bm{\nabla} \theta) = 0,
\end{align}
which is the bosonic continuity equation. The real part is given by
\begin{align}\label{eq:RePart_GPE}
     \partial_t \theta = -\frac{1}{2} (\bm{\nabla} \theta)^2 -g \rho - Q(\rho)+\mu
\end{align}
with the quantum potential $Q(f) = - (\nabla^2\sqrt{f})/ 2\sqrt{f}$ as defined in the main text. We split the phase variable into a smooth and a vortex part, $\theta = \theta_\mathrm{sm} + \theta_v$. In order to obtain an equation for the velocity field $\bm v$, we take the gradient of~\eqref{eq:RePart_GPE}. With 
\begin{align}
    \bm{\nabla} (\partial_t \theta) &= \bm{\nabla} (\partial_t \theta_\mathrm{sm} + \partial_t \theta_v) = \partial_t \bm{\nabla} (\theta -\theta_v) +\bm{\nabla} \partial_t \theta_v \nonumber\\
    &= \partial_t \bm v + (\bm{\nabla} \partial_t - \partial_t \bm{\nabla}) \theta_v
\end{align}
and the definition of the vortex 3-current (see~\eqref{eq:Def_vortex_current}) as
\begin{align}
    j_\mu^v = \frac{1}{2\pi} \epsilon_{\mu\nu\lambda} \partial_\nu\partial_\lambda \theta_v,
\end{align}
we obtain 
\begin{align}
    \partial_t \bm{v}+ \frac{1}{2}\bm{\nabla}  v^2= -\nabla \left( Q(\rho) + g \rho \right) - 2\pi \, \hat{z}\times \bm j_v.
\end{align}
The vortex current term is hence identified as
\begin{align}
(\bm\nabla\partial_t-\partial_t\bm\nabla)\theta_v =2\pi\,\hat{\bm z}\times\bm j_v.
\end{align}
Using the identity
\begin{align}
    \frac{1}{2}\bm{\nabla}  v^2 =( \bm v \cdot \bm{\nabla} ) \,  \bm v + \bm v \times (\bm{\nabla} \times \bm v)
\end{align}
and the definition of the vortex density 
\begin{align}
    \rho_v = j_0^v = \frac{1}{2\pi} \epsilon_{ij} \partial_i \partial_j \theta_v = \frac{1}{2\pi} \epsilon_{ij} \partial_i \partial_j \theta = \frac{1}{2\pi} (\bm{\nabla} \times \bm v)_z, \nonumber
\end{align}
we obtain
\begin{align}
    \left( \partial_t + \bm v \cdot \bm{\nabla} \right) \bm v = -\bm{\nabla} \left( Q(\rho) + g \rho \right) + \bm f_\mathrm{Lorentz},
\end{align}
with $\bm f_\mathrm{Lorentz} = 2\pi \, \hat{z} \times (\rho_v \bm v - \bm j_v)$ as quoted in the main text (Eq.~\eqref{eq:Hydrodynamic_GPE}). The vortex-current contribution records the motion of the phase singularities, specifying their trajectories completes the hydrodynamic description.

\section{Guiding center motion of vortices in boson coordinates}\label{app:Guiding_center}
In this appendix, we present an alternate derivation of the vortex velocity starting from the superfluid theory in boson coordinates. This may be directly obtained from the GP equation, but we start here from the superfluid action~\cite{ao1993berry,thouless1996transverse}. 

In the lowest Landau level limit (guiding center approximation), the Lorentz force acting on vortex $a$ may be written as
\begin{align}
    \label{eq:Lorentz_force_balance}
    \bm{F}_L &= \, -\left[ \bm e(\bm X_a) + \dot{\bm X_a}\times b(\bm X_a)\,\hat{ z} \right]\nonumber \\
    &=  -2\pi \, \rho(\bm X_a)\, \hat{ z}\times \left[ \bm v(\bm X_a) -\dot{\bm X}_a \right] = 0,
\end{align}
where the vortices are treated as having no inertia, since there is no additional kinetic term for the vortices present in the Lagrangian~\eqref{eq:Dual_Lagrangian}. From the Lorentz force balance~\eqref{eq:Lorentz_force_balance}, we obtain the vortex velocity 
\begin{equation}
    \dot{\bm{X}}_a = \bm{v}(\bm{X}_a),
\end{equation}
which is consistent with Eq.~\eqref{eq:vortex-velocity} derived from the dual equations of motion in Appendix~\ref{app:Dual_theory}. In this Appendix, we shall derive this equation in an equivalent manner to the previous formulation using the worldline action, except make manifest that the gauge field $a_\mu$ may be interpreted as a Berry connection in configuration space of the vortices. In this formulation, $a_\mu $ (denoted by $A_\mu$ in this appendix) is a static field unlike in the dual theory.   

We wrote the superfluid Lagrangian density in density and phase variables in Eq.~\eqref{eq:lsf_density_phase_variables}. As it turns out, the term that is responsible for the Lorentz force equation (vortex velocity) is the Berry term,
\begin{equation}
    L_{\rm Berry} = - \int \dd^2 x \,  \rho\,\partial_t\theta.
\end{equation}
Consider a vortex at position $\bm X(t)$ with unit charge. We decompose the phase as before $ \theta(\bm x,t) = \theta_v(\bm x, t) + \theta_\mathrm{sm}(\bm x,t)$ as in Appendix~\ref{app:Hydrodynamics_from_GPE}. Note that we can parametrize the vortex part of the phase as $\theta_v = \theta_v(\bm{x}-\mathbf{X}(t))$, where
\begin{align}
    &\theta_v(\bm x-\bm X) = \arg(\bm x-\bm X), \nonumber\\ &\nabla \times \nabla \theta_v = 2\pi \,\delta^{(2)}(\bm x-\bm X)\,\hat{\bm z}
\end{align}
Inserting this back into the Berry term, we obtain
\begin{align}
    &L_{\rm Berry} = A_{i}(\bm X,t) \, \dot X_{i}, \nonumber \\
    &A_{i}(\bm X,t) \equiv - \int \dd^2 x \, \rho(\bm x,t)\, \partial_{X_i}\theta_v(\bm x-\bm X).
    \label{eq:Berry_connection}
\end{align}
Notice that the derivatives are taken with respect to the position of the vortex $\{X_i\}$. In fact, $A_i$ can be interpreted as a Berry connection, which lies in vortex configuration space. The corresponding Berry curvature is
\begin{align}\label{eq:definition_Berry_curvature}
    F_{ij} = \partial_{X_{i}}A_j(\bm X,t)-\partial_{X_{j}}A_i(\bm X,t).
\end{align}
Simplifying further, we obtain,
\begin{align}
    F_{ij} &=-\int \textrm{d}^2x\,\rho(\bm x,t)\left(\partial_i\partial_j-\partial_j\partial_i\right)\theta_v \nonumber \\
    &= -2\pi \,\rho(\bm X,t)\,\epsilon_{ij}
    \label{eq:evaluation_Berry_curvature}
\end{align}
To elucidate this geometric structure further, one can examine the action corresponding to the Berry term,
\begin{align}
    S_{\rm Berry}  = \int \dd t \, L_{\rm Berry} = \int \dd t \, \, A_{i}(\bm X) \, \dot X_{i}. 
\end{align}
Now consider a closed vortex trajectory $\mathcal{C}$ of a single vortex at position $\bm X$, with
$\bm X(T)=\bm X(0)$. The Berry action is
\begin{align}
    S_{\textrm{Berry}}[\mathcal{C}] =\int_0^T \dd t\, A_i(\bm X)\dot X_i = \oint_{\mathcal{C}} A_i(\bm X)\, \dd X_i ,
\end{align}
Here, we assume that the background density is held fixed while the vortex is moved around the loop. By Stokes' theorem,
\begin{align}
    S_{\textrm{Berry}}[\mathcal{C}] &=\int_{\Sigma(\mathcal{C})} \dd^2 X\,\epsilon_{ij} \partial_{X_i} A_j  = \int_{\Sigma(\mathcal{C})} \dd^2 X\, \frac{1}{2} \epsilon_{ij} F_{ij}  \nonumber\\
    &= - 2 \pi \int_{\Sigma(\mathcal{C})} \dd^2 X\, \rho(\bm X)=  - 2 \pi   \, \rho_s \, \Sigma(\mathcal{C}), 
\end{align}
where $\Sigma(\mathcal{C})$ is the area enclosed by the vortex path and we inserted the Berry curvature $F_{ij}$ from~\eqref{eq:definition_Berry_curvature} and~\eqref{eq:evaluation_Berry_curvature}. The same term can be understood as the usual Berry connection associated with a family of vortex states parametrized by the vortex coordinate $\bm X$. In quantum mechanics, a family of states
$\ket{\Psi(\bm X)}$ has Berry connection
\begin{align}
     A_i(\bm X) = \textrm{i}\mel{\Psi(\bm X)}{\partial_{X_i}}{\Psi(\bm X)} .
\end{align}
For a condensate vortex state, schematically,
\begin{align}
    \Psi_{\bm X} \sim \prod_{\alpha=1}^N e^{\textrm{i}\theta_v(\bm r_\alpha-\bm X)}\times \text{regular part}
\end{align}
assuming an $N$ boson state. Then, evaluating the Berry connection yields,
\begin{align}
    A_i(\bm X)   &=-\sum_\alpha \, \mel*{\Psi(\bm X)}{\partial_{X_i} \theta_v(\bm r_\alpha-\bm X(t))}{\Psi(\bm X)}\nonumber\\
    &= \int \dd^2 x \, \rho(\bm x)\, \partial_i\theta_v(\bm x-\bm X(t)).
\end{align}
This reproduces the Berry connection~\eqref{eq:Berry_connection}. The vortex thus moves in an effective magnetic field proportional to the condensate density. For a closed path $\mathcal{C}$, the Berry phase is the effective flux through the area enclosed by the vortex trajectory,
\begin{align}
    S_{\rm Berry}[\mathcal{C}] = -2\pi \int_{\Sigma(\mathcal{C})} \dd^2X\,\rho(\bm X).
\end{align}
Defining $ N_{\rm enc} = \int_{\Sigma(C)} \dd^2X\,\rho(\bm X)$ as the number of bosons enclosed by the vortex loop, the Berry phase is $2\pi N_{\rm enc}$. The condensate many-body wavefunction accumulates the phase $e^{-\textrm{i}2\pi q N_{\rm enc}}$. This may be interpreted as the bosons acting as magnetic flux quanta for the vortices. In the macroscopic GP description the density is a smooth field, so this phase varies continuously with the area enclosed by the vortex path. The corresponding local Berry curvature is responsible for the transverse Lorentz force.

To obtain the full force acting on a vortex, we must also examine the vortex ``potential energy". For a single vortex we define the scalar potential ($\hbar=m=1$),
\begin{align}
\label{eq:def_scalar_potential}
    U(\{\bm X\},t) = \int \textrm{d}^2x\, \left[ \frac{\rho}{2}(\nabla\theta)^2 \right]_{\theta= \theta_v(\bm x-\bm X)+\theta_\mathrm{sm}} .
\end{align}
For simplicity, we set $\theta_{\textrm{sm}}=0$ ahead in this section. Then, $(\rho/2)(\bm{\nabla}\theta)^2 = (\rho/2) \sum_j (\partial_j \theta_v(\bm x - \bm X))^2$. To the Berry term in the Lagrangian we then add the scalar potential, so that the vortex Lagrangian is
\begin{equation}
    L_{\textrm{vortex}}=  \bm A(\bm X,t) \,  \dot{\bm X }_a(t) - U(\{ \bm X\},t).
\end{equation}
The Euler-Lagrange equation for each vortex yields,
\begin{equation}
    \frac{\dd A_i(\bm X, t)}{\dd t} - ( \partial_{X_{i}} A_j)\, \dot X_{j} + \partial_{X_{i}} U(\{\bm X \}, t) = 0. 
\end{equation}
Using $\frac{\dd A_i}{\dd t} = \partial_t A_i+ (\partial_{X_{j}} A_i) \, \dot{X}_{j} $, this can be written as
\begin{align}   
    \partial_t A_i - F_{ij} \, \dot X_{j} + \partial_{X_{i}} U(\{\bm X\}, t) = 0.
    \label{eq:Magnus_force_components}
\end{align}
We shall now evaluate the first and last term in this equation explicitly using the expressions for the gauge potential and the scalar potential. We first find,
\begin{align}
    \partial_t A_i  &=  \int \dd^2 x \ \partial_t \rho(\bm x,t)\, \partial_i\theta_v(\bm x-\bm X) \nonumber \\
    &= -\int \dd^2 x \ \partial_j J_j(\bm x,t)\, \partial_i\theta_v(\bm x-\bm X) \nonumber \\
    &= \int \dd^2 x \ J_j(\bm x,t)\, \partial_j \partial_i\theta_v(\bm x-\bm X_a)  \text{ + boundary terms} \nonumber
\end{align}
We also evaluate the derivative on the $U(\{\bm X,t\})$ appearing in~\eqref{eq:Magnus_force_components}. For this we use~\eqref{eq:def_scalar_potential} and obtain
\begin{align}
     \partial_{X_{i}} U(\{\bm X \}, t)   &=\frac{1}{2} \partial_{X_{i}} \bigg[\int \dd^2 x \  \rho(\bm x, t) \sum_j \big(\partial_j \theta_v(\bm x - \bm X)\big)^2\bigg] \nonumber\\
     &=- \int \dd^2 x \, \rho(\bm x, t) \, v_j\,\partial_i \partial_j\theta_v(\bm x-\bm X(t)) . \nonumber
\end{align}
Putting both terms in~\eqref{eq:Magnus_force_components} together, we obtain
\begin{align}
    \partial_t A_i &+ \partial_{X_{i}} U(\{\bm X \}, t)  \nonumber \\
    &= - \int \dd^2 x \, \rho \, v_j\,\left(\partial_i \partial_j - \partial_j \partial_i\right)\theta_v(\bm x-\bm X(t)) \nonumber\\
    &= - \int \dd^2 x \, \rho \, v_j\, 2\pi  \, \epsilon_{ij} \, \delta^{(2)} (x-\bm X). \nonumber \\
    &=  - 2\pi  \, \epsilon_{ij} \,\rho(\bm X, t) \, v_j  (\bm X, t)
\end{align}
Finally using the Berry curvature expression from~\eqref{eq:evaluation_Berry_curvature}, we find the Lorentz force acting on each vortex is (from~\eqref{eq:Magnus_force_components}),
\begin{align}
    \bm F_{\rm L} = - 2\pi \, \rho(\bm X_a)\, \hat{ z}\times \left[ \bm v(\bm X) -\dot{\bm X} \right] = 0.
\end{align} 
This matches the Lorentz force~\eqref{eq:Lorentz_force_balance} acting on the vortex charges obtained in the dual coordinates. It corresponds to the lowest Landau level limit for the vortices. 

\section{Mutual friction and dissipative current relaxation}\label{app:mutual_friction}

We derive the phenomenological vortex drift law in Eq.~\eqref{eq:vortex_drift_law} of the main text from a balance of Lorentz and drag forces, and relate it to dissipative vortex transport and relaxation of the boson current. We assume a finite temperature superfluid where the normal fluid component represents thermal excitations in Landau's two-fluid picture. Its velocity is prescribed rather than evolved, and its microscopic interactions with vortices are summarized by friction coefficients~\cite{skaugen2017origin,mehdi2023mutual}. 

\subsection{Vortex drift law}

Consider a vortex $a$ with circulation $2\pi q_a$ with vortex charge $q_a=\pm1$. Let $\bm \dot{X}_a$ be its velocity, $\bm v_s$ the regular superfluid velocity evaluated at its position, and $\bm v_n$ the normal-fluid velocity there. The vortex's singular self velocity is excluded from the point-vortex force balance. We introduce the operator $\mathsf R$ for a counterclockwise rotation by $90^\circ$ in the plane,
\begin{equation}
 \mathsf R\bm a=\hat{\bm z}\times\bm a,\qquad
 \mathsf R^2=-\mathbbm{1},\qquad
 \bm a\cdot\mathsf R\bm a=0.
 \label{eq:mf-rotation}
\end{equation}
In a minimal model with scalar drag coefficient $D\geq0$, the forces on the vortex are~\cite{hall1956rotation}
\begin{align}
 \bm F_L&=q_a\, G\,\mathsf R(\bm \dot{X}_a-\bm v_{s,a}),\\
 \bm F_D&=-D(\bm \dot{X}_a-\bm v_n).
 \label{eq:mf-forces}
\end{align}
In the continuum local-drift description, the Lorentz force is evaluated at the position of the vortex given by $\bm v_{s,a}=\bm v(\bm X_a)$, with the regular velocity understood. The coefficient $G$ in the Lorentz force is given by the magnetic field $G=b\geq0$ according to Eq.~\eqref{eq:Lorentz_force}. The drag $F_D$ opposes motion relative to the normal component, while the Lorentz force $F_L$ is transverse to motion relative to the superfluid. With no vortex inertia in the guiding center approximation (lowest Landau level limit), the force balance reads
\begin{equation}
 \bm F_L+\bm F_D=0.
 \label{eq:mf-force-balance}
\end{equation}
Define
\begin{equation}
 \bm U=\bm \dot{X}_a-\bm v_{s,a},\quad
 \bm W=\bm v_n-\bm v_{s,a},\quad
 \beta=\frac{D}{G}.
 \label{eq:mf-relative-velocities}
\end{equation}
Equation~\eqref{eq:mf-force-balance} becomes
\begin{equation}
 q_a\mathsf R\bm U-\beta(\bm U-\bm W)=0,
 \quad
 (\beta\mathbbm{1}-q_a\mathsf R)\bm U=\beta\bm W.
\end{equation}
Since $q_a^2=1$, the inverse operator follows from
\begin{equation}
 (\beta\mathbbm{1}+q_a\mathsf R)
 (\beta\mathbbm{1}-q_a\mathsf R)
 =(1+\beta^2)\mathbbm{1}.
\end{equation}
The exact solution of this scalar-drag model is therefore
\begin{align}
 \bm U
 &=\frac{\beta^2}{1+\beta^2}\bm W
   +\frac{\beta}{1+\beta^2}q_a\mathsf R\bm W,\\
 \bm \dot{X}_a
 &=\bm v_s+\alpha q_a\mathsf R(\bm v_n-\bm v_{s,a})
              +\alpha'(\bm v_n-\bm v_{s,a}), \label{eq:mf-general-drift}\\
 \alpha&=\frac{\beta}{1+\beta^2},\qquad
 \alpha'=\frac{\beta^2}{1+\beta^2}.
 \label{eq:mf-drag-coefficients}
\end{align}
The relation between $\alpha$ and $\alpha'$ in Eq.~\eqref{eq:mf-drag-coefficients} is specific to the minimal scalar-drag model. Additional transverse forces change this relation. The $\alpha'$ term is reactive in the fluid energy balance.

In the weak-drag limit $D \ll G$,
\begin{equation}
 \alpha=\beta+\mathcal{O}(\beta^3),\qquad
 \alpha'=\beta^2+\mathcal{O}(\beta^4),
\end{equation}
and hence
\begin{equation}
 \bm \dot{X}_a=\bm v_{s,a}+\beta q_a\mathsf R(\bm v_n-\bm v_{s,a}) +\mathcal{O}(\beta^2|\bm v_n-\bm v_{s,a}|).
\end{equation}
For a stationary normal component, $\bm v_n=0$, Eq.~\eqref{eq:mf-general-drift} reduces to
\begin{equation}
 \bm \dot{X}_a=(1-\alpha')\bm v_{s,a}-\alpha q_a\mathsf R\bm v_{s,a}.
\end{equation}
In our numerical simulations, we retain the dissipative coefficient and set $\alpha'=0$, resulting in
\begin{equation}
 \dot{\bm X}_a
 =\bm v_{s,a}-\alpha q_a\hat{\bm z}\times\bm v_{s,a}.
 \label{eq:mf-production-drift}
\end{equation}
Neglecting the reactive term is hence consistent to first order in $\beta$ in the weak-drag expansion.

\subsection{Dissipative vortex transport}

The general drift law~\eqref{eq:mf-general-drift} yields
\begin{equation}
\begin{aligned}
 \bm j_v&=\rho_v\bm v+\bm j_v^{\mathrm{diss}}+\bm j_v^{\mathrm{react}},\\
 \bm j_v^{\mathrm{diss}}
 &=\alpha n_v\mathsf R(\bm v_n-\bm v),\\
 \bm j_v^{\mathrm{react}} &= \alpha'\rho_v(\bm v_n-\bm v).
 \label{eq:mf-current-general}
\end{aligned}
\end{equation}
The dissipative term involves $n_v=\sum_a\delta(\bm x-\bm X_a(t))$, rather than $\rho_v$, because multiplying the frictional drift by the transported charge produces $q_a^2$. Vortices and antivortices therefore contribute with the same sign to dissipative transport. The reactive term retains one factor of the signed charge.

We set $\alpha'=0$ as used for our numerical simulations and define the dual field relative to the normal component,
\begin{align}
 \bm{\mathcal E}_n
 &\equiv\bm e-b\mathsf R\bm v_n
       =b\mathsf R(\bm v-\bm v_n),\\
 \sigma_v&\equiv\frac{\alpha n_v}{b}\geq0.
 \label{eq:mf-effective-electric-field}
\end{align}
Since $\mathsf R\bm v=\bm e/b$, the circulation current becomes
\begin{equation}
 \bm j_v^{\mathrm{diss}}=-\sigma_v\bm{\mathcal E}_n.
 \label{eq:mf-conductivity-circulation}
\end{equation}
For $\bm v_n=0$, the effective field is simply $\bm{\mathcal E}_n=\bm e$. Eq.~\eqref{eq:mf-conductivity-circulation} lends itself to an Ohmic-like interpretation of mutual friction in dual variables.

\subsection{Relaxation of the boson current}

From Eq.~\eqref{eq:mf-current-general}, we obtain the explicit frictional acceleration,
\begin{align}
 \left.\partial_t\bm v\right|_{\mathrm{diss}}
 &=-2\pi\mathsf R\bm j_v^{\mathrm{diss}}\nonumber \\
 &=2\pi\alpha n_v(\bm v_n-\bm v)
   -2\pi\alpha'\rho_v\mathsf R(\bm v_n-\bm v).
 \label{eq:mf-acceleration-general}
\end{align}
For $\alpha'=0$ this can also be obtained directly from the conductivity law:
\begin{align}
 \left.\partial_t\bm v\right|_{\mathrm{diss}}=2\pi\sigma_v\mathsf R\bm{\mathcal E}_n=-2\pi\alpha n_v(\bm v-\bm v_n).
 \label{eq:mf-acceleration-ohmic}
\end{align}
The notation for an explicit frictional contribution therefore means
\begin{equation}
 \left.\partial_t\bm J\right|_{\mathrm{diss}}
 =-2\pi\alpha n_v(\bm J-\rho\bm v_n),
 \qquad 
\end{equation}
or
\begin{equation} 
 \left.\partial_t\bm J\right|_\mathrm{diss}
 =-2\pi\alpha n_v\bm J \text{ for } \bm v_n = 0. 
 \label{eq:mf-boson-current-relaxation}
\end{equation}
The continuity equation has no separate mutual-friction source at fixed instantaneous fields. Nevertheless, friction changes subsequent density evolution through its effect on $\bm v$. The Ohmic-like vortex response and boson-current relaxation are consequently two descriptions of the same term in the coupled equations.

\section{Numerical algorithm to evolve dual equations of motion} \label{app:numerical_algorithm}

To numerically evolve the equations of motion, we use a coupled field-particle framework with a particle-mesh discretization of the classical hydrodynamic equations. The fields $b$ and $\bm w = \bm e/b$ are represented on a periodic spatial grid, while vortices carry quantized charges and move in continuous, real-valued coordinates. This discretization treats vortices
as classical particles coupled to grid fields, rather than as
matter fields on a lattice. Spatial derivatives are evaluated in Fourier space by multiplying each mode by its corresponding wavevector component. The vortex trajectories supply a vortex current (entering the equations of motion) through deposition of the vortex positions on the numerical grid for the fields. On the other hand, interpolation evaluates the grid fields at the vortex position needed to update vortex velocities.  At each timestep, guessed vortex trajectories supply an initial current, for which the field equations are solved iteratively to prescribed convergence tolerances. The resulting fields determine updated vortex trajectories and hence a revised current. The algorithm alternates field solves and trajectory updates updates until both satisfy the convergence criteria, after which the timestep is accepted.

\subsection{Numerical method}
The parameters used in the numerical simulations (except for the separate dipole calculation in Appendix~\ref{app:dipole_scaling}) are
\begin{equation}
 \begin{gathered}
 L_x=L_y=L=40,\qquad N_x=N_y=N=128,\\ h=L/N=0.3125,
 \Delta t=10^{-3},\qquad m=1, \\\qquad b_0=500,\qquad g=6.285,
 \end{gathered}
 \label{eq:run-parameters}
\end{equation}
with periodic boundary conditions, where $N$ is the number of grid points along each direction $x$ and $y$. Thus $g\approx500$. The simulation spans $6\times10^6$ timesteps, corresponding to $t=6000$. The initial fields are $b(\bm x,0)=b_0$ and $\bm v(\bm x,0)=0$, and there are no vortices at $t=0$. With these parameters, we obtain the healing length as
\begin{equation}
  \xi = \frac{1}{\sqrt{2m g \rho_0}} \approx 0.0316
\end{equation}
and hence $L \approx 1265\xi$.

\subsubsection{Continuum equations and regularizations}

The evolved grid variables are the dual magnetic field $b$ and the velocity-like field $\bm w=\bm e/b$. The fields $b$ and $\bm w$ are discretized on the given grid. The vortex sector consists of positions $\bm X_a$ on the torus and fixed integer charges $q_a=\pm1$, where $a=1, \dots, N$ labels the vortices. With $\bm v=\bm w\times \hat{\bm z}=(w_y,-w_x)$, the equations evolved by the algorithm are
\begin{align}
 \partial_t b&=-\curl(b\bm w) =-\div(b\bm v), \label{eq:b-eom}\\
 \partial_t\bm w &=-\frac{1}{m}\hat{\bm z}\times\nabla {\cal H}_b +\frac{2\pi}{m}\bm j_v-\gamma\bm w^T, \label{eq:u-eom}\\
 \div\bm w&=-\frac{2\pi}{m}\rho_v, \label{eq:gauss-law}\\
 \rho_v(\bm x,t)&=\sum_a q_a S(\bm x-\bm X_a),\\ 
 \bm j_v(\bm x,t)&=\sum_aq_a\dot{\bm X}_aS(\bm x-\bm X_a), \label{eq:vortex-density-current}
\end{align}
where $S(\bm x)$ is a shape function with finite width that converges to the expected Delta-function for point particles in the limit $h \rightarrow 0$. Here $\bm w^T$ is the divergence-free Fourier component of $\bm w$, $\div \bm w^T = 0$. The Hamiltonian derivative appearing in Eq.~\eqref{eq:u-eom} is
\begin{align}
 {\cal H}_b
 &=\frac{m}{2}|\bm w|^2+\frac{g}{2\pi}b+Q(b),
 \label{eq:Hb}\\
 Q(b)&=-\frac{1}{2m}\frac{\nabla^2\sqrt{b}}{\sqrt{b}},
 \label{eq:quantum-pressure}
\end{align}
The quantum pressure is evaluated using $b$ directly. The kinetic Hamiltonian is quadratic in $\bm w$.

All spatial derivatives use Fourier pseudospectral operators, e.g. $\partial_i f={\cal F}^{-1}(ik_i\widetilde f)$ and $\nabla^2f={\cal F}^{-1}(-k^2\widetilde f)$ with $\widetilde f$ being the Fourier transform of $f$. Nonlinear products are evaluated on the original real-space grid, without padding. For real fields on an even grid, the derivative in each coordinate has zero contribution from its corresponding Nyquist mode. Neutral pair creation and annihilation maintain $\sum_aq_a=0$, as required by the periodic Gauss' law.

\subsubsection{Spectral vortex density and charge-conserving current}
\label{sec:cic}

In contrast to the fields $\bm w$ and $b$ that are defined on the discrete grid, vortex positions are real-valued coordinates on the periodic domain. The coupled solver uses a translation-invariant spectral cloud with the cloud-in-cell (CIC) hat-function form factor~\cite{hockney1988computer}, rather than depositing onto the four nearest nodes.

The form factor follows from the normalized one-dimensional CIC cloud,
\begin{equation}
 s_h(x)=\begin{cases}
 h^{-1}(1-|x|/h),& |x|\leq h,\\
 0,& |x|>h,
 \end{cases}
\end{equation}
which, in standard nodal CIC deposition, assigns the vortex charge to the two grid points bracketing its position with linear weights. Here, we instead use its Fourier transform to define a translation-invariant spectral cloud.
Since $s_h(x)$ is real and even, its Fourier transform is given by
\begin{align}
 \nonumber \widehat s_h(k)=&\frac{2}{h}\int_0^h\!\mathrm{d}x\,
       (1-x/h)\cos(kx)
 =\\
 &\frac{2[1-\cos(kh)]}{(kh)^2}
 =\left[\frac{\sin(kh/2)}{kh/2}\right]^2,
\end{align}
with $\widehat s_h(0)=1$ by continuity. The two-dimensional cloud is the product $S(x,y)=s_h(x)s_h(y)$, periodically continued over the simulation domain. Its Fourier coefficients therefore factorize as $\widehat S(\bm k)=\widehat s_h(k_x)\widehat s_h(k_y)$ at the allowed periodic wavevectors. Here, $\widehat S(\bm k)$ denotes the continuous Fourier transform of the normalized cloud shape, evaluated at wavevector $\bm k$, so that $\widehat S(\bm 0)=1$. In contrast, a symbol $\widetilde{.}$ denotes the unnormalized discrete Fourier transform of a grid field, such as the vortex density $\widetilde\rho_{v,\bm k}$ in $k$-space below. The conversion between these conventions results in an additional factor $N^2/L^2$ in Eq.~\eqref{eq:cic-density}.

Using $\operatorname{sinc}z=\sin z/z$, we obtain
\begin{equation}
 \widehat S(\bm k)=\operatorname{sinc}^2(k_xh/2)
                       \operatorname{sinc}^2(k_yh/2).
 \label{eq:cic-weights}
\end{equation}
With the unnormalized discrete Fourier transform on the $N\times N$ grid, the deposited vortex density on the discrete grid is
\begin{equation}
 \widetilde\rho_{v,\bm k}=\frac{N^2}{L^2}\widehat S(\bm k)
       \sum_aq_a e^{-i\bm k\cdot\bm X_a}.
 \label{eq:cic-density}
\end{equation}
The zero mode of the charge density $(\sum_a q_a = 0)$ and both Nyquist lines are removed from $ \widetilde\rho_{v,\bm k}$. In the following, we denote the retained modes by $\mathcal R$. Vortex current deposition uses the same Nyquist projection but retains its uniform mode, since there is no sum rule for the uniform mode of vortex current in contrast to vortex density. To evaluate grid fields at vortex positions, we use an interpolation operator given by the adjoint of deposition with respect to the grid quadrature,
\begin{equation}
 \mathcal I_a[f]=\frac{1}{N^2}\sum_{\bm k\in\mathcal R}
       \widehat S(\bm k)\widetilde f_{\bm k}
       e^{i\bm k\cdot\bm X_a}.
\end{equation}
The interpolation sum includes the $\bm k=0$ mode, so that spatially uniform fields are retained. For direct cloud interpolation, the contribution of a vortex's own longitudinal Gauss-law field vanishes by symmetry. Deposition and interpolation supply opposite position phases and the same real, even form factor, resulting in a self-field proportional to
\begin{equation}
 \sum_{\substack{\bm k\in\mathcal R\\\bm k\neq0}}
 \frac{i\bm k}{k^2}\widehat S(\bm k)^2=0.
\end{equation}
The retained modes occur in pairs $\bm k$ and $-\bm k$, whose contributions cancel to roundoff, independently of the vortex position relative to the grid.

For an unwrapped displacement $\Delta\bm X_a$ over a step, the exact straight-path current on the resolved modes is~\cite{villasenor1992rigorous,esirkepov2001exact}
\begin{align}
 \widetilde{\bm J}_{v,\bm k}
 &=\frac{N^2}{L^2}\widehat S(\bm k)\sum_aq_a
     \frac{\Delta\bm X_a}{\Delta t}
     e^{-i\bm k\cdot(\bm X_a^n+\Delta\bm X_a/2)}\notag\\
 &\hspace{2cm}\times\operatorname{sinc}(\bm k\cdot\Delta\bm X_a/2).
 \label{eq:spectral-path-current}
\end{align}
It satisfies $
\frac{\widetilde\rho_{v,\bm k}^{n+1}-\widetilde\rho_{v,\bm k}^{n}}{\Delta t}
+i\bm k\cdot\widetilde{\bm J}_{v,\bm k}=0 $ to roundoff, including paths crossing the periodic boundary. Thus the field update preserves Gauss' law to the nonlinear solve tolerance without an ordinary-step projection.

\subsubsection{Guiding-center drift motion}
Including mutual friction with $\alpha=5\times10^{-4}$ and zero reactive friction, the vortex velocity is implemented as
\begin{align}
 \bm v_a\equiv\dot{\bm X}_a
 &=\bm v_{s,a}-\alpha \, q_a\, \hat{\bm z}\times\bm v_{s,a}.
 \label{eq:mutual-friction}
\end{align}
Here $\bm v_{s,a}$ denotes the guiding-center velocity without mutual friction. In the continuum description, it is the local superfluid velocity at the vortex position,
\begin{equation}
 \bm v_{s,a}=\bm v(\bm X_a)
 =-\hat{\bm z}\times\bm w(\bm X_a).
\end{equation}
For a vortex represented by a finite cloud, we outline two options of evaluating this velocity from the grid fields.

The first option is direct cloud interpolation,
\begin{equation}
 \bm v_{s,a}=-\hat{\bm z}\times\mathcal I_a[\bm w].
\end{equation}
This is the more immediate discretization of the continuum guiding-center law. It eliminates the contribution of a vortex's own longitudinal Gauss-law field because $\mathcal I_a[\bm w_{\rm self}]=0$ by the Fourier-mode symmetry cancellation described above. However, when $b$ varies across the cloud, this prescription generally does not make the advective vortex-current work vanish for the existing field-particle coupling.

The second option, used in the numerical implementation, is the density-weighted cloud velocity,
\begin{equation}
 \bm v_{s,a}=-\hat{\bm z}\times
       \frac{\mathcal I_a[b\bm w]}{\mathcal I_a[b]}.
 \label{eq:vortex-drift}
\end{equation}
This choice ensures an energy-conserving implementation of the guiding center drift given the Hamiltonian. The numerator in~\eqref{eq:vortex-drift} averages the mass current across the vortex cloud, while the denominator normalizes by the corresponding averaged density, giving greater weight to regions containing more fluid. With this choice, advective vortex motion does no net work on the fields. Since interpolation is adjoint to current deposition, the instantaneous advective power is
\begin{equation}
 P_{j_v}^{\rm adv}
 =\int\mathrm{d}^2x\,b\bm w\cdot\bm j_v^{\rm adv}
 =\sum_a q_a\,\mathcal I_a[b\bm w]\cdot\bm v_{s,a}=0.
\end{equation}
The last equality follows because the drift is perpendicular to $\mathcal I_a[b\bm w]$. The density-weighted prescription hence preserves the corresponding discrete energy balance. Note that this zero-power identity concerns only the conservative guiding-center contribution, not the added mutual friction dissipation.

The two options thus involve a tradeoff at finite cloud width. Direct interpolation preserves cancellation of the longitudinal self-field contribution, whereas density weighting preserves the exact zero-power identity. For nonuniform $b$, the latter can retain a self contribution because $\mathcal I_a[b\bm w_{\rm self}]$ need not vanish: density variations weight opposite sides of the circulating self flow differently. This contribution should not be identified a priori as a numerical error. Coupling to an inhomogeneous density may have physical content, since density and ambient phase gradients contribute to vortex motion in the Gross-Pitaevskii theory~\cite{groszek2018motion}. Preserving the discrete energy balance provides a conservative treatment of the finite cloud, but neither conservation nor this physical analogy establishes that the weighted self contribution reproduces the correct vortex drift. That requires comparison with resolved vortex-core dynamics.

For uniform $b$, the two prescriptions coincide and both the self-field cancellation and zero-power identity hold. They also agree in the smooth-field continuum limit: as the cloud width $h\to0$, with $b$ and $\bm w$ regular near $\bm X_a$ and $b(\bm X_a)>0$, both approach $-\hat{\bm z}\times\bm w(\bm X_a)$. This argument does not establish equivalence in a singular point-vortex limit or at a core where the density vanishes.

For a finite timestep, the trial vortex displacement defines a straight path from the old position to the trial new position. We average the cloud interpolation operator along this entire path, rather than evaluating it at a single vortex position. In Eq.~\eqref{eq:vortex-drift}, this path average is applied separately to the numerator and denominator, using the arithmetic endpoint averages $\bar b=(b^n+b^{n+1})/2$ and $\bar{\bm w}=(\bm w^n+\bm w^{n+1})/2$ to form $\bar b\,\bar{\bm w}$ and $\bar b$. Using the same path for interpolation and current deposition preserves their adjoint relation and makes the advective vortex-current work vanish at convergence.

\subsubsection{Coupled field update}

\begin{figure*}[!t]
   \centering
    \includegraphics[width=\linewidth]{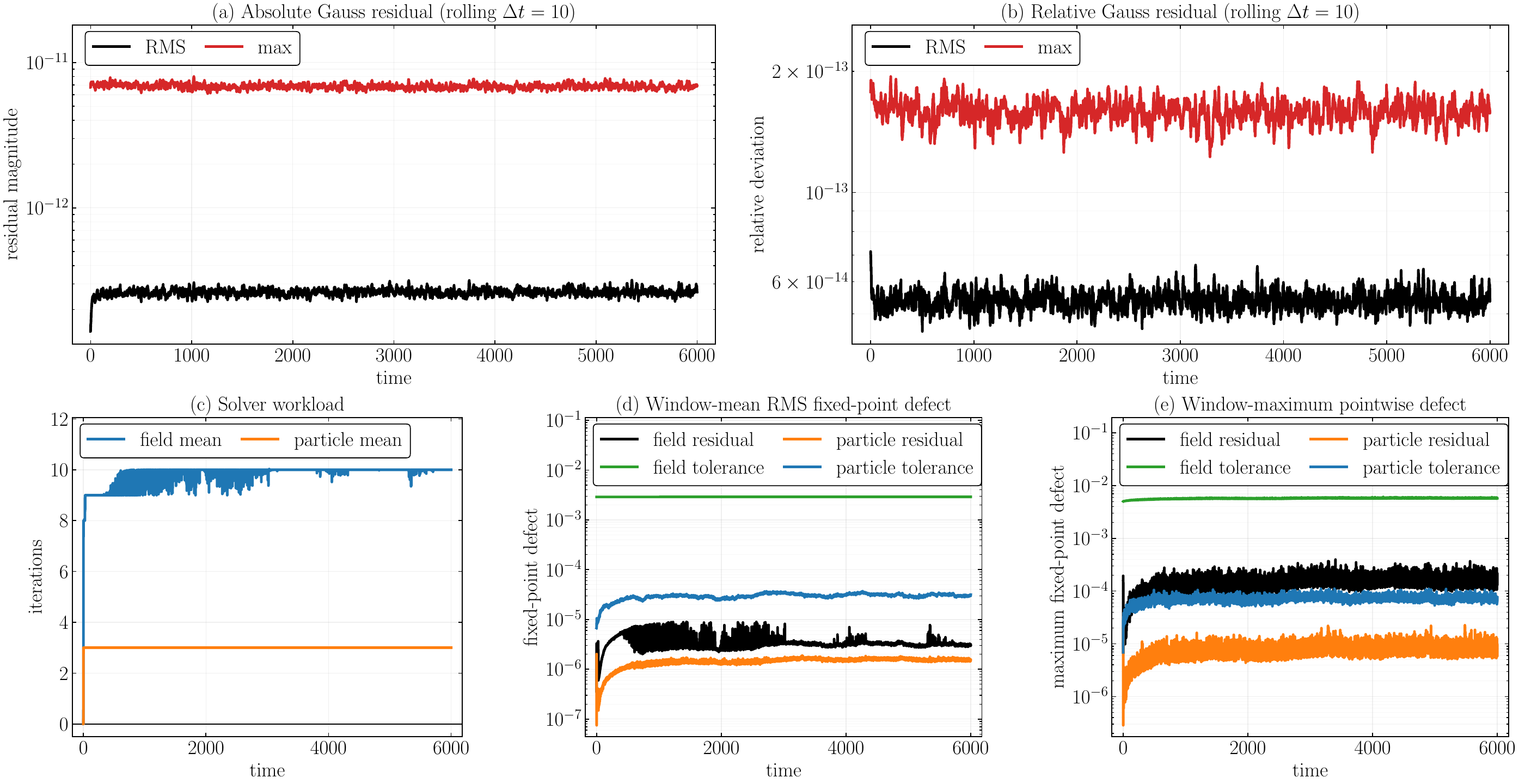}
    \caption{Gauss law constraint and convergence of the coupled field-particle solver. \textbf{(a,b)} RMS and maximum absolute and relative Gauss-law residuals, respectively, averaged within the rolling window indicated in the titles. \textbf{(c)} Mean iteration counts for the field and particle solves in each diagnostics window. \textbf{(d)} Window-mean RMS and \textbf{(e)} window-maximum pointwise fixed-point defects for both solves, together with their corresponding acceptance tolerances.}
    \label{fig:S1}
\end{figure*}

We advance the conservative field equations with an implicit, symmetric discrete-gradient integrator~\cite{gonzales1996time}. For each trial vortex trajectory, the associated charge-conserving current is held fixed while the magnetic and velocity fields are solved together for their new endpoints. This field solve is nested within the trajectory iteration, so the fields and vortex motion are mutually consistent when the timestep converges. The discrete Hamiltonian gradient ensures that the conservative exchange terms cancel in the grid energy balance, up to the nonlinear solve tolerance.

The integrator stores $b^n$, $\bm w^n$ and $\bm X_a^n$ at the same integer time. For a fixed trajectory current, it solves the endpoints simultaneously:
\begin{subequations}
\begin{align}
 \frac{b^{n+1}-b^n}{\Delta t}
   &=-\curl(\bar b\,\bar{\bm w}), \label{eq:implicit-kick1}\\
 \frac{\bm w^{n+1}-\bm w^n}{\Delta t}
   &=-\frac{1}{m}\hat{\bm z}\times\nabla\overline{\mathcal H}_b
       +\frac{2\pi}{m}\bm J_v, \label{eq:implicit-kick2}\\
 \bar b=\tfrac12(b^n+&b^{n+1}),\quad
 \bar{\bm w}=\tfrac12(\bm w^n+\bm w^{n+1}). \label{eq:implicit-kick3}
\end{align}
\end{subequations}
The force uses a two-point discrete Hamiltonian gradient rather than simply evaluating Eq.~\eqref{eq:Hb} at the midpoint. In terms of the fields at times $t_n$ and $t_{n+1}$, it is
\begin{align}
 \overline{\mathcal H}_b
 &=\frac{m}{4}(|\bm w^n|^2+|\bm w^{n+1}|^2)
       +\frac{g}{4\pi}(b^n+b^{n+1})+\overline Q,\\
 \overline Q&=\frac{1}{8m}\left[
   -\frac{|\nabla b^n|^2+|\nabla b^{n+1}|^2}{2b^nb^{n+1}}\right.\notag\\
 &\left.\quad-\div\left\{\frac{\nabla b^n+\nabla b^{n+1}}{2}
                   \left(\frac{1}{b^n}+\frac{1}{b^{n+1}}\right)\right\}\right].
 \label{eq:discrete-quantum-gradient}
\end{align}
The same spectral gradient and its negative adjoint divergence are used throughout. This discrete gradient gives the exact change of the grid Hamiltonian under the symmetric endpoint product rule. At a converged conservative step, the field exchange terms cancel algebraically. The adjoint, density-weighted path law also makes the advective vortex-current work vanish; finite solver tolerances leave a residual energy error.

An outer Picard iteration starts from an instantaneous particle-velocity predictor. Each trial displacement defines $\bm J_v$. An inner Picard solve advances both fields, and the resulting midpoint fields define a corrected path velocity. Subsequent field solves start from the preceding field endpoint. Both the field and particle fixed points must converge before the step is accepted. The accepted displacement is wrapped onto the torus only when storing the new position.

For each field block $f=b,\bm w$ independently, the unrelaxed fixed-point defect must satisfy
\begin{align}
 r_{f,\rm rms}&\leq10^{-6}+10^{-5}s_{f,\rm rms},\\
 r_{f,\max}&\leq10^{-7}+10^{-5}s_{f,\max}.
\end{align}
Here $s_f$ is the candidate field scale, and vector norms use the full two-component magnitude. The field solve permits 2-30 inner iterations with unit relaxation. Particle defects are displacement changes divided by $\Delta t$. Their RMS and maximum thresholds are respectively $10^{-8}+10^{-5}s_{v,\rm rms}$ and $10^{-8}+10^{-5}s_{v,\max}$, with at most ten outer iterations. These are global RMS and maximum criteria, whereas per-particle residual ratios are diagnostic only. A nonpositive or nonfinite path-averaged interpolated density also rejects the particle map. Failed attempts are discarded and the interval is recursively split into two half steps. At depth $d$, the field iteration limit is multiplied by $2^d$, with a maximum subdivision depth of ten. Exhausted retries reject the nominal step.

In \cref{fig:S1}(a,b), we plot the Gauss' law absolute and relative residuals, which are of the order of $10^{-11}$ and $10^{-13}$, respectively. From that, we conclude that Gauss' law is fulfilled up to a very high precision. \Cref{fig:S1}(c) shows the field and particle iteration counts to be approximately 10 and 5 respectively, showcasing that the solver workload is manageable which means the algorithm is converging as expected and no hard solve steps are encountered. In \cref{fig:S1}(d,e), we plot the fixed-point defects relative to their acceptance tolerances. We notice that both the field and particle residual lie well below their tolerance values, indicating rapid convergence of the iterative solution at each timestep.  

After the coupled solve, spectral damping with $\gamma=3\times10^{-3}$ acts only on the divergence-free component,
\begin{equation}
 \widetilde{\bm w}^{T}\longrightarrow
       e^{-\gamma\Delta t}\widetilde{\bm w}^{T}.
\end{equation}
This leaves the longitudinal Gauss-law field unchanged. Damping and density repair are accounted for as separate energy changes, followed by the vortex events described below.

\subsection{Vortex pair production and annihilation}
\label{sec:vortex-events}

Vortex creation and annihilation are applied after each successful ordinary field/particle step. Vortex pair production is implemented as a homogeneous Poisson process of rate $\Gamma=10$: At every step, the number of generated vortex pairs is 
\begin{equation}
  N_{\rm pair}\sim {\rm Poisson}(\Gamma\Delta t). 
\end{equation}
Here $\Gamma\Delta t=0.01$ is the expected number of injected dipoles per step. More than one event in the same step is permitted. For each event a midpoint $\bm X_c$ is sampled uniformly on the torus and an orientation $\theta$ chosen uniformly on $[0,2\pi)$. A neutral dipole of fixed separation $d_f=1.5$ is inserted at
\begin{equation}
\begin{aligned}
 \bm X_+&=\bm X_c+\frac{d_f}2 (\cos\theta,\sin\theta),\\ \bm X_-&=\bm X_c-\frac{d_f}2 (\cos\theta,\sin\theta), 
 \end{aligned}
\end{equation}
with both positions wrapped periodically. Proposed positions are resampled until each new vortex is at least $h=0.3125$ from all existing vortices and from vortices already inserted during the same step. Thus every event adds one $q=+1$ and one $q=-1$ vortex and leaves $\sum_aq_a=0$. Importantly, pair creation leaves $b$ unchanged, such that $\rho_v$ is redeposited and only $\bm w = \bm e /b$ is minimally projected by adding a longitudinal gradient so that the Gauss law~\eqref{eq:gauss-law} holds for the new particle list. This instantaneous projection supplies the velocity field of the newly created pair and its associated energy is attributed to the pair-production event.

Annihilation is checked every step, after all pair production steps. All opposite-sign pairs whose minimum-image torus distance is strictly smaller than $d_a=1.0$ are collected and sorted by increasing separation. After annihilation, $\rho_v$ is redeposited and $\bm w$ is projected longitudinally back onto Gauss' law. The dipole separation at production exceeds the annihilation radius, although a newly created vortex may annihilate immediately with a closer pre-existing opposite charged vortex.

\begin{figure*}[!t]
   \centering
    \includegraphics[width=\linewidth]{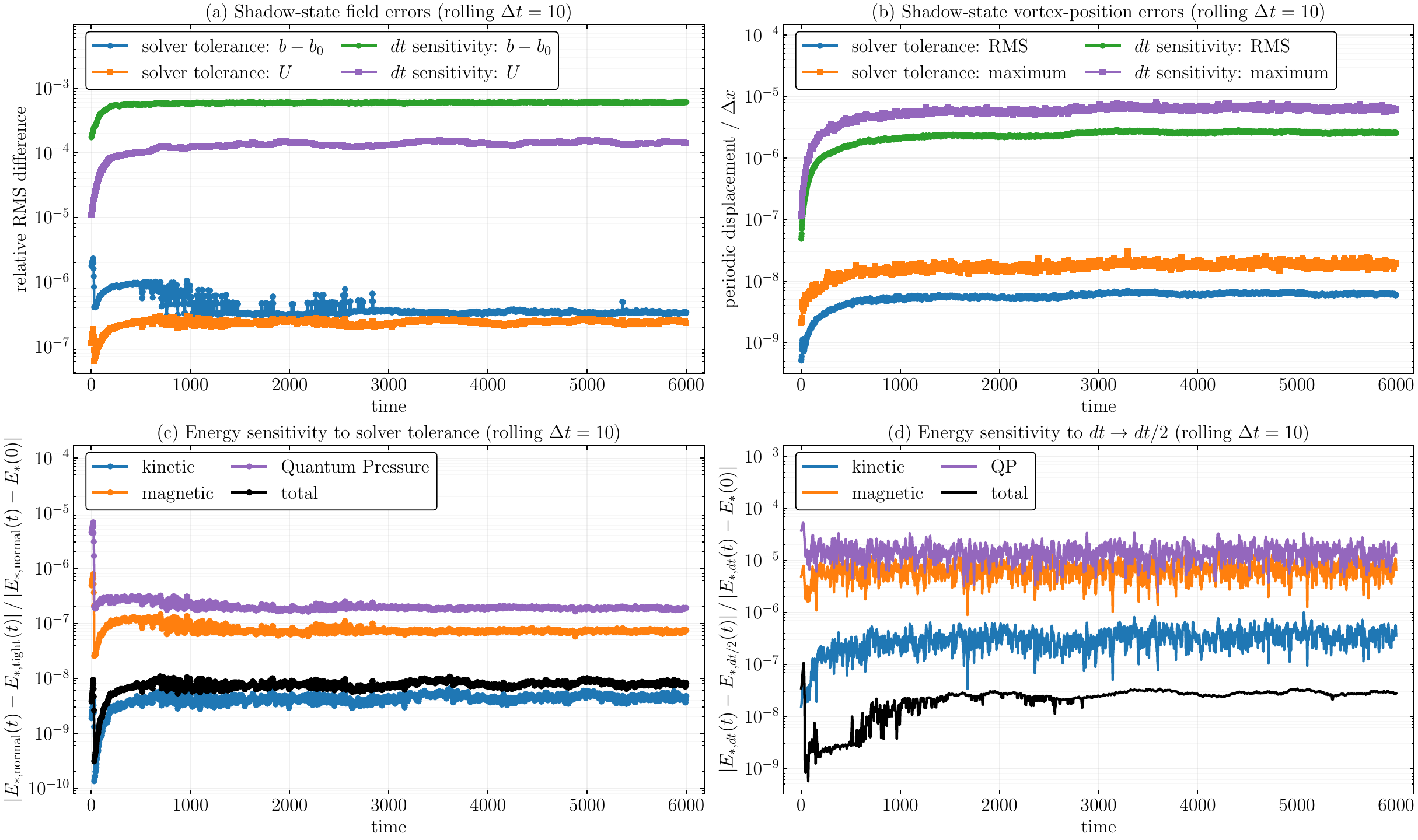}
    \caption{Numerical sensitivity of algorithm though shadow evolutions starting from the same production state. \textbf{(a)} Relative RMS differences in the magnetic-density fluctuations and density-weighted velocity, and \textbf{(b)} RMS and maximum vortex-position differences in units of the grid spacing. Both panels compare tighter solver tolerances and two half steps at normal tolerance with the production full step. \textbf{(c)} Sensitivity of the kinetic, magnetic, quantum-pressure and total energies to tightening the solver tolerances. \textbf{(d)} Energy sensitivity to replacing one full step by two half steps at normal tolerance. The energy differences are normalized by the corresponding candidate's accumulated energy change, $|E_{*,\mathrm{candidate}}(t)-E_*(0)|$, rather than its absolute background energy. All curves use the rolling windows indicated in the titles.}
    \label{fig:S2}
\end{figure*}

\subsection{Evaluation of the tracked energies}
\label{sec:energies}

Observables are evaluated at integer times.
For the coupled integrator the stored fields are already simultaneous,
\begin{equation}
 \bm w^n_{\rm diag}=\bm w^n.
 \label{eq:diagnostic-velocity}
\end{equation}
The reported electric field is $\bm e^n=b^n\bm w^n_{\rm diag}$. The three Hamiltonian contributions are
\begin{align}
 E_{\rm kin}&=\frac{m}{4\pi}\int\dd^2x\,
 b|\bm w_{\rm diag}|^2,\label{eq:Ekin}\\
 E_b&=\frac{g}{8\pi^2}\int\dd^2x\,b^2,\label{eq:Eb}\\
 E_Q&=\frac{1}{16\pi m}\int\dd^2x\,
 \frac{|\nabla b|^2}{b},\label{eq:EQ}\\
 E_{\rm tot}&=E_{\rm kin}+E_b+E_Q.
 \label{eq:Etotal}
\end{align}
Every integral is the rectangular sum $h^2\sum_{IJ}$. Equation~\eqref{eq:Eb} includes the uniform background contribution $gL_xL_yb_0^2/(8\pi^2)$. 

For scale-resolved kinetic diagnostics, we form the density-weighted velocity
\begin{equation}
 \bm W(\bm x)=\sqrt{\frac{mb(\bm x)}{2\pi}}\, \bm w_{\rm diag}(\bm x), \ E_{\rm kin}=\frac12\int\dd^2x\,|\bm W(\bm x)|^2,
 \label{eq:density-weighted-velocity}
\end{equation}
and apply the Fourier Helmholtz projectors
\begin{equation}
 P^L_{ij}(\bm k)=\frac{k_ik_j}{k^2},\qquad  P^T_{ij}(\bm k)=\delta_{ij}-\frac{k_ik_j}{k^2}
 \label{eq:helmholtz}
\end{equation}
to the Fourier transformed $\widetilde{\bm W}_{\bm k}$. Using the unnormalized discrete Fourier transform, the energy assigned to a Fourier mode is
\begin{equation}
 \varepsilon^{A}_{\bm k}=\frac{h^2}{2N^2}  \left|P^A(\bm k)\widetilde{\bm W}_{\bm k}\right|^2, \qquad A=L,T. \label{eq:modal-energy}
\end{equation}
Modes are binned into shells (label $s$) of width $\Delta k=0.1$ and radius $k_s$. Then $E_A(k_s)=\Delta k^{-1}\sum_{\bm k\in s}\varepsilon^A_{\bm k}$, such that $\sum_sE_A(k_s)\Delta k=E_{\rm{kin}, A}$ for $A=L,T$, where $E_{\rm{kin}, A}$ is the total kinetic energy for the incompressible ($A=L$) and compressible ($A=T$) part. The $\bm k=0$ contribution is included in the total kinetic spectrum but in neither projected spectrum. 

To evaluate instantaneous energy-transfer rates at $t_n$, we construct $\bm j_v^n$ by depositing each vortex's charge times its instantaneous velocity, $q_a\bm v_a^n$, using the spectral cloud centered at $\bm X_a^n$. This diagnostic current differs from the timestep-averaged path current used to evolve the fields and enforce discrete charge conservation. The associated instantaneous power is monitored as
\begin{equation}
 P_{j_v}=\int\dd^2x\,b\ \bm w_{\rm diag}\cdot\bm j_v.
 \label{eq:vortex-current-power}
\end{equation}
It vanishes in the continuum for pure guiding-center advection. With the matched density-weighted cloud law, the instantaneous advective contribution vanishes by the discrete grid-particle work identity. The dissipative mutual friction can give a nonzero contribution. Solver residuals, density repair, damping and vortex events are tracked separately in the energy budget. Scalar energies and spectra are averaged over diagnostic windows, with sampling every 10 nominal steps and saving every 100 nominal steps. The saved vortex count is the instantaneous integer count at the endpoint of that averaging window.

\subsection{Numerical error estimates}

We estimate local numerical errors by recomputing selected accepted steps with discarded shadow states. The shadow evolutions are computed every 2000 nominal steps. All shadow evolutions start from the same production state at $t_n$ and end at $t_n+\Delta t$, before vortex creation or annihilation. They include the ordinary field and particle evolution and the damping and density repair. The audit does not change the production trajectory or its random-number streams.

Let $\Phi_{\Delta t}^{\tau}$ denote the ordinary step map with the production collection of field and particle tolerances $\tau$. In addition to the accepted production endpoint $Z_{\rm p}$, we compute two shadow endpoints,
\begin{align}
 Z_{\rm t}&=\Phi_{\Delta t}^{\tau/100}(Z^n),\\
 Z_{\rm h}&=\Phi_{\Delta t/2}^{\tau}\circ
              \Phi_{\Delta t/2}^{\tau}(Z^n).
 \label{eq:error-audit-shadows}
\end{align}
Here $Z$ comprises both grid fields and the vortex positions. Tightening divides all absolute and relative RMS and maximum tolerances by 100, for both the inner field solve and the outer particle corrector. To permit convergence, the tight shadows increase both the field iteration limit and the particle iteration limit by a factor of four. Interval subdivision is disabled for these tight solves, so an accepted tight shadow retains the requested step size. The production-tolerance half steps retain the ordinary retry policy, and the audit records whether the production full step used retries. Comparisons requiring a failed shadow are left undefined rather than treating an unconverged endpoint as a reference.

We compare the ordered candidate-reference pairs $(Z_{\rm p},Z_{\rm t})$ and $(Z_{\rm p},Z_{\rm h})$. The first probes the error from terminating the nonlinear solves at the production tolerances. The second measures sensitivity to halving the timestep under the production solver settings. They measure local sensitivity on the fixed spatial grid, rather than accumulated trajectory error or spatial convergence.

For each comparison, with candidate $C$ and reference $R$, the relative field errors are
\begin{align}
 \epsilon_b&=\frac{\|b_C-b_R\|_{\rm rms}}
   {\max(\|b_R-b_0\|_{\rm rms},\epsilon_{\rm machine}\cdot \max(|b_0|,1))},\\
 \epsilon_W&=\frac{\|\bm W_C-\bm W_R\|_{\rm rms}}
   {\max(\|\bm W_R\|_{\rm rms},\epsilon_{\rm machine})},
 \label{eq:error-audit-field-metrics}
\end{align}
where $\bm W=\sqrt{mb/(2\pi)}\,\bm w$ as in Eq.~\eqref{eq:density-weighted-velocity}, and $\epsilon_{\rm machine}$ is machine precision. Taking the maximum with $\epsilon_{\rm machine}$ prevents division by zero or a vanishingly small denominator. The RMS is taken over all grid entries, including both components for $\bm W$. Subtracting $b_0$ in the density reference scale prevents the uniform background from masking errors in density fluctuations. Vortices are matched by their particle identifiers. Their minimum-image displacement differences on the torus are reported as RMS and maximum distances in units of $h$. If the identifier lists do not match, these position errors are undefined. Empty matching lists have zero position error.

The audit also evaluates $E_{\rm kin}$, $E_b$, $E_Q$ and $E_{\rm tot}$ at both endpoints and stores their absolute differences, candidate values and relative differences $|E_C-E_R|/\max(|E_R|,\epsilon_{\rm machine})$. The error estimates are evaluated at the audited endpoint and retained only when the corresponding shadow solves converge. These shadow-state estimates supplement the fixed-point residuals and energy-budget diagnostics, which alone do not determine the error of the evolved state.

\Cref{fig:S2} shows the resulting field, vortex-position and energy sensitivities. In \cref{fig:S2}(c,d), we normalize the energy difference between shadow evolutions by the candidate's accumulated energy change $|E_{*,\mathrm{candidate}}(t)-E_*(0)|$. This emphasizes sensitivity relative to the evolving energy rather than the uniform background contribution. Panel (d) shows the timestep comparison at the production solver tolerances. All shadow-field errors and relative energy sensitivity measures are small and constant over time. From that, we conclude that the employed timestep and tolerance values are sufficiently small for the algorithm to converge to a numerical solution at each timestep that is accurate within the determined error bounds.

\section{Kinetic energy spectrum and flux analysis for compressible superfluids}\label{app:fluxes}

We compute the flux decomposition of the density weighted velocity~\cite{jha2025energy} $\bm u = \sqrt{\rho} \,\bm v$ in the following.

\subsection{Compressible and incompressible spectra}
Following the main text, we decompose $\bm u$ as 
\begin{align}
  \bm u=\bm u_c+\bm u_i, \qquad   \curl\bm u_c=0, \qquad   \div\bm u_i=0
\end{align}
into the incompressible ($\bm u_i$) and compressible ($\bm u_c$) component. For $\bm k\ne0$, the corresponding projectors to the compressible ($P_c$) and incompressible ($P_i$) part in Fourier space are
\begin{align}
  [P_c]_{\alpha\beta}(\bm k) &=\frac{k_\alpha k_\beta}{k^2}, &   [P_i]_{\alpha\beta}(\bm k)  &=\delta_{\alpha\beta}-\frac{k_\alpha k_\beta}{k^2}, \nonumber \\
  [u_a]_\alpha(\bm k) &=[P_a]_{\alpha\beta}(\bm k)u_\beta(\bm k), &a&=c,i.\label{eq:flux-physical-projectors}
\end{align}
Using the Fourier transformation
\begin{align}
  \bm u(\bm x,t)
  =\int\frac{\dd^2k}{(2\pi)^2}\,
  \bm u(\bm k,t)\ e^{i\bm k\cdot\bm x},
\end{align}
we define the kinetic energy spectrum~\cite{nore1997kolmogorov,kobayashi2005kolmogorov-2}
\begin{align}
  E_a(k,t)
  \equiv
  \frac{1}{(2\pi)^2}\int\dd\Omega_k^{(2)}\,k\,
  \frac12\left|\bm u_a(\bm k,t)\right|^2
  \label{eq:flux-physical-spectrum}
\end{align}
for the two components $a=c,i$. The total kinetic energy is then given by
\begin{align}
  E_{\rm kin}(t)
  =\int_0^\infty\dd k\,[E_c(k,t)+E_i(k,t)].
\end{align}
The time derivative of the kinetic energy spectra are 
\begin{align}
  \partial_tE_a(k,t)
  =\mathcal \int_\Omega\operatorname{Re}\left[
  \bm u_a^*(\bm k,t)\cdot\partial_t\bm u(\bm k,t)
  \right]
  \label{eq:flux-spectrum-time-derivative}
\end{align}
for $a=i,c$, where, for compactness, we introduced the shell integral
\begin{align}
  \mathcal \int_\Omega 
  \equiv\frac{1}{(2\pi)^2}
  \int\dd\Omega_k^{(2)}\,k.
  \label{eq:flux-shell-operator}
\end{align}

\subsection{Equation of motion}
The equation of motion of the velocity $\bm v$ is 
\begin{align}
  \partial_t\bm v =-\nabla \left[\frac{\bm v^2}{2}+g\rho+Q(\rho)\right] -2\pi\,\bm \hat{\bm z}\times j_v-\gamma\bm v_c,
  \label{eq:flux-physical-velocity-eom}
\end{align}
with the boson continuity equation
\begin{align}
  \partial_t \rho + \div (\rho \bm v) = 0
  \label{eq:flux-continuity}
\end{align}
and the compressible part of the superfluid velocity $\bm v_c$. 
Away from instantaneous pair-production and annihilation events, the vortex velocity including mutual friction is
\begin{align}
  \dot{\bm X}_a =\bm v(\bm X_a)-\alpha\, q_a\, \hat{\bm z}\times\bm v(\bm X_a). \label{eq:flux-mutual-friction-velocity}
\end{align}
It is useful to introduce the signed and unsigned vortex densities
\begin{align}
  \rho_v(\bm x,t)&=\sum_aq_a\delta(\bm x-\bm X_a), & n_v(\bm x,t)&=\sum_a\delta(\bm x-\bm X_a).
\end{align}
In the continuum point-vortex description, the vortex current then separates into guiding-center and mutual-friction parts,
\begin{align}
  \bm j_v =\rho_v\bm v-\alpha \,n_v\,\hat{\bm z}\times\bm v. \label{eq:flux-vortex-current-split}
\end{align}
Consequently,
\begin{align}
  \sqrt\rho\,\hat{\bm z} \times \bm j_v =\rho_v\,\hat{\bm z}\times \bm u +\alpha \,n_v\,\bm u. \label{eq:flux-vortex-source-split}
\end{align}
By taking the time derivative of $\bm u=\sqrt\rho\bm v$, using
Eq.~\eqref{eq:flux-continuity}, and expressing $\bm v$ through $\bm u$, we obtain the equation of motion for $\bm u$,
\begin{align}
 \nonumber  \partial_t\bm u
  ={}&\bm R_{\rm kin}+\bm R_{\mathrm{g}}+\bm R_{\mathrm{Q}}+\bm R_\rho
  +\bm R_{\mathrm{v}}-\bm R_\alpha-\bm R_\gamma
  +\\
  &\bm R_{\rm pair}-\bm R_{\rm ann},
  \label{eq:flux-density-weighted-eom}
\end{align}
where the individual contributions are
\begin{align*}
  \bm R_{\rm kin} &\equiv-\sqrt\rho\,\nabla\left(\frac{\bm u^2}{2\rho}\right),\\
  \bm R_{\mathrm{g}} &\equiv-g\sqrt\rho\,\nabla\rho,\\
  \bm R_{\mathrm{Q}} &\equiv-\sqrt\rho\,\nabla Q(\rho),\\
  \bm R_\rho &\equiv-\frac{\bm u}{2\rho}\,
  \div(\sqrt\rho\,\bm u),\\
  \bm R_{\mathrm{v}} &\equiv -2\pi\,\rho_v\,\hat{\bm z}\times \bm u,\\
  \bm R_\alpha &\equiv 2\pi\, \alpha \, n_v\,\bm u,\\
  \bm R_\gamma &\equiv\gamma\sqrt\rho\,\bm v_c.
\end{align*}
With the Fourier convention above, we denote
\begin{align*}
  \mathcal F[f](\bm k)
  &\equiv\int\dd^2x\,f(\bm x)e^{-i\bm k\cdot\bm x},
  &
  \int_{\bm p}
  &\equiv\int\frac{\dd^2p}{(2\pi)^2},
\end{align*}
and write $s\equiv\sqrt\rho$, $r\equiv\rho^{-1}$, and $s^{-1}\equiv1/\sqrt\rho$. The Fourier transformation of the individual terms in Eq.~\eqref{eq:flux-density-weighted-eom} are then
\begin{align*}
  \mathcal F[\bm R_{\rm kin}](\bm k)  &=-\frac{i}{2}\int_{\bm p}\int_{\bm q}\int_{\bm \ell} s(\bm k-\bm p)\,\bm p\,r(\bm p-\bm q)\times  \\
  &\hspace{10pt}\left[\bm u(\bm q-\bm \ell)\cdot\bm u(\bm \ell)\right],\\
  \mathcal F[\bm R_{\mathrm{g}}](\bm k) &=-ig\int_{\bm p}s(\bm k-\bm p)\,\bm p\,\rho(\bm p),\\
  \mathcal F[\bm R_{\mathrm{Q}}](\bm k) &=-i\int_{\bm p}s(\bm k-\bm p)\,\bm p\,Q(\bm p), \\
  Q(\bm p) &=\frac12\int_{\bm q}s^{-1}(\bm p-\bm q)\,q^2s(\bm q),\\
  \mathcal F[\bm R_\rho](\bm k) &=-\frac{i}{2}\int_{\bm p}\int_{\bm q}\int_{\bm \ell} \bm u(\bm k-\bm p)\,r(\bm p-\bm q)\times \, \\
  &\hspace{10pt}\bm q\cdot\left[s(\bm q-\bm \ell)\bm u(\bm \ell)\right],\\
  \mathcal F[\bm R_{\mathrm{v}}](\bm k) &=-2\pi\int_{\bm p}\rho_v(\bm k-\bm p) \left[\hat{\bm z}\times\bm u(\bm p)\right],\\
  \mathcal F[\bm R_\alpha](\bm k) &=2\pi\alpha\int_{\bm p}n_v(\bm k-\bm p)\bm u(\bm p),\\
  \mathcal F[\bm R_\gamma](\bm k) &=\gamma\int_{\bm p}s(\bm k-\bm p)\bm v_c(\bm p).
\end{align*}

\subsection{Shell transfers and cumulative fluxes}

We characterize energy redistribution using shell-resolved transfer rates and their cumulative fluxes, following the standard spectral energy-budget framework~\cite{verma2019energy}. For each nondissipative term $\chi\in\{\mathrm{kin},\mathrm{g},\mathrm{Q},\rho,\mathrm{v}\}$, we define the shell-resolved kinetic transfer
\begin{align}
  T_\chi^a(k,t)
  \equiv\mathcal \int_\Omega\operatorname{Re}\left[
  \bm u_a^*(\bm k,t)\cdot
  \mathcal F[\bm R_\chi](\bm k,t)
  \right],
  \qquad a=c,i.
  \label{eq:flux-shell-transfers}
\end{align}
Explicitly, all of the evaluated transfer channels are therefore
\begin{align}
  \nonumber T_{\rm kin}^a &=\frac12\mathcal \int_\Omega\operatorname{Im}\bigg[ \bm u_a^*(\bm k)\cdot \int_{\bm p}\int_{\bm q}\int_{\bm \ell} s(\bm k-\bm p)\,\bm p\,r(\bm p-\bm q) \times \\
  &\hspace{10pt}\left(\bm u(\bm q-\bm \ell)\cdot\bm u(\bm \ell)\right)\bigg],
  \label{eq:flux-transfer-kin}
\end{align}
\begin{align}
  T_{\mathrm{g}}^a &=g\mathcal \int_\Omega\operatorname{Im}\left[\bm u_a^*(\bm k)\cdot \int_{\bm p}s(\bm k-\bm p)\,\bm p\,\rho(\bm p)\right],
  \label{eq:flux-transfer-g}
\end{align}
\begin{align}
  \nonumber T_{\mathrm{Q}}^a &=\frac12\mathcal \int_\Omega\operatorname{Im}\bigg[ \bm u_a^*(\bm k)\cdot  \int_{\bm p}\int_{\bm q}s(\bm k-\bm p)\,\bm p\times \\
  & \hspace{10pt}s^{-1}(\bm p-\bm q)\,q^2s(\bm q) \bigg],
  \label{eq:flux-transfer-Q}
\end{align}
\begin{align}
  \nonumber T_\rho^a &=\frac12\mathcal \int_\Omega\operatorname{Im}\bigg[  \bm u_a^*(\bm k)\cdot  \int_{\bm p}\int_{\bm q}\int_{\bm \ell}  \bm u(\bm k-\bm p)\,r(\bm p-\bm q)\times \\  
  &\hspace{10pt}\bm q\cdot\left(s(\bm q-\bm \ell)\bm u(\bm \ell)\right)\bigg],
  \label{eq:flux-transfer-density-weighting}
\end{align}
\begin{align}
  T_{\mathrm{v}}^a &= - 2\pi\mathcal \int_\Omega\operatorname{Re}\left[ \bm u_a^*(\bm k)\cdot \int_{\bm p}\rho_v(\bm k-\bm p) \left(\hat{\bm z}\times\bm u(\bm p)\right)\right].
  \label{eq:flux-transfer-vortex-current}
\end{align}
The Bernoulli / phonon-work channel (label $\mathrm{p}$) combines the Bernoulli and density-weighting contributions. The advective vortex-current channel is labeled $\mathrm{v}$, while $\mathrm{g}$ and $\mathrm{Q}$ denote the magnetic and quantum-pressure channels, respectively.
\begin{align}
  T_{\mathrm{p}}^a\equiv T_{\rm kin}^a+T_\rho^a,\qquad \bm R_{\mathrm{p}}\equiv\bm R_{\rm kin}+\bm R_\rho.
\label{eq:flux-phonon-channel}
\end{align}
The mutual-friction dissipation term is
\begin{align}
  D_\alpha^a(k,t) \equiv2\pi\alpha\mathcal \int_\Omega\operatorname{Re}\left[   \bm u_a^*(\bm k,t)\cdot \mathcal F[n_v\bm u](\bm k,t) \right].
  \label{eq:flux-mutual-friction-dissipation}
\end{align}
Although its projected shell contributions need not be positive separately, the total mutual-friction power is nonnegative:
\begin{align}
  \sum_{a=c,i}\int_0^\infty\dd k\,D_\alpha^a(k,t) =2\pi\alpha\int\dd^2x\,n_v(\bm x,t)\bm u^2(\bm x,t)\ge0.
  \label{eq:flux-mutual-friction-positive}
\end{align}
The compressible-velocity damping contributes the shell rate
\begin{align}
  D_\gamma^a(k,t)\equiv\gamma\mathcal \int_\Omega\operatorname{Re}\left[\bm u_a^*(\bm k,t)\cdot\mathcal F[\sqrt\rho\,\bm v_c](\bm k,t)\right],
  \hspace{10pt} a=c,i.
  \label{eq:flux-velocity-damping}
\end{align}
Here $\bm v_c$ is the compressible part of $\bm v$, rather than $\bm u_c/\sqrt\rho$. For nonuniform density, $\sqrt\rho\,\bm v_c$ can have both Helmholtz components, so $D_\gamma^i$ need not vanish and $D_\gamma^c$ is not generally equal to $2\gamma E_c$. The sign convention assigns positive $D_\gamma^a$ to energy loss; individual projected shell rates can have either sign.
We denote the shell-local forcing from pair production by $F_{\rm pair}^a(k,t)$ and the shell-local loss from annihilation by $D_{\rm ann}^a(k,t)$, without attempting to express either event term as a smooth function of the pre-event fields. The kinetic spectral balance is then
\begin{align}
  \nonumber \partial_tE_a(k,t)
  =&\sum_{\chi\in\{\mathrm{p},\mathrm{g},\mathrm{Q},\mathrm{v}\}}
  T_\chi^a(k,t)
  -D_\alpha^a(k,t)-D_\gamma^a(k,t)
  +\\
  &F_{\rm pair}^a(k,t)-D_{\rm ann}^a(k,t),
  \qquad a=c,i.
  \label{eq:flux-kinetic-shell-balance}
\end{align}
Finally, the cumulative transfer flux associated with each evaluate $T$-channel is~\cite{numasato2010direct,billam2015spectral}
\begin{align}
  \Pi_\chi^a(k,t)
  \equiv-\int_0^k\dd k'\,T_\chi^a(k',t),
  \qquad
  \chi\in\{\mathrm{p},\mathrm{g},\mathrm{Q},\mathrm{v}\}.
  \label{eq:flux-cumulative-definition}
\end{align}
Thus $\partial_k\Pi_\chi^a=-T_\chi^a$, and positive $\Pi_\chi^a(k)$ includes transfer from modes below $k$ to modes above $k$ as well as energy conversion to a different energy component. With
\begin{align}
  \Pi_a \equiv\Pi_{\mathrm{p}}^a+\Pi_{\mathrm{g}}^a+\Pi_{\mathrm{Q}}^a+\Pi_{\mathrm{v}}^a,
\end{align}
the balance can equivalently be written
\begin{align}
  \nonumber &\partial_t E_a(k,t)+\partial_k\Pi_a(k,t)
  =\\
  &F_{\rm pair}^a(k,t)-D_\alpha^a(k,t)-D_\gamma^a(k,t)-D_{\rm ann}^a(k,t).
  \label{eq:flux-kinetic-balance-cumulative}
\end{align}

\begin{figure*}[!t]
   \centering
    \includegraphics[width=\linewidth]{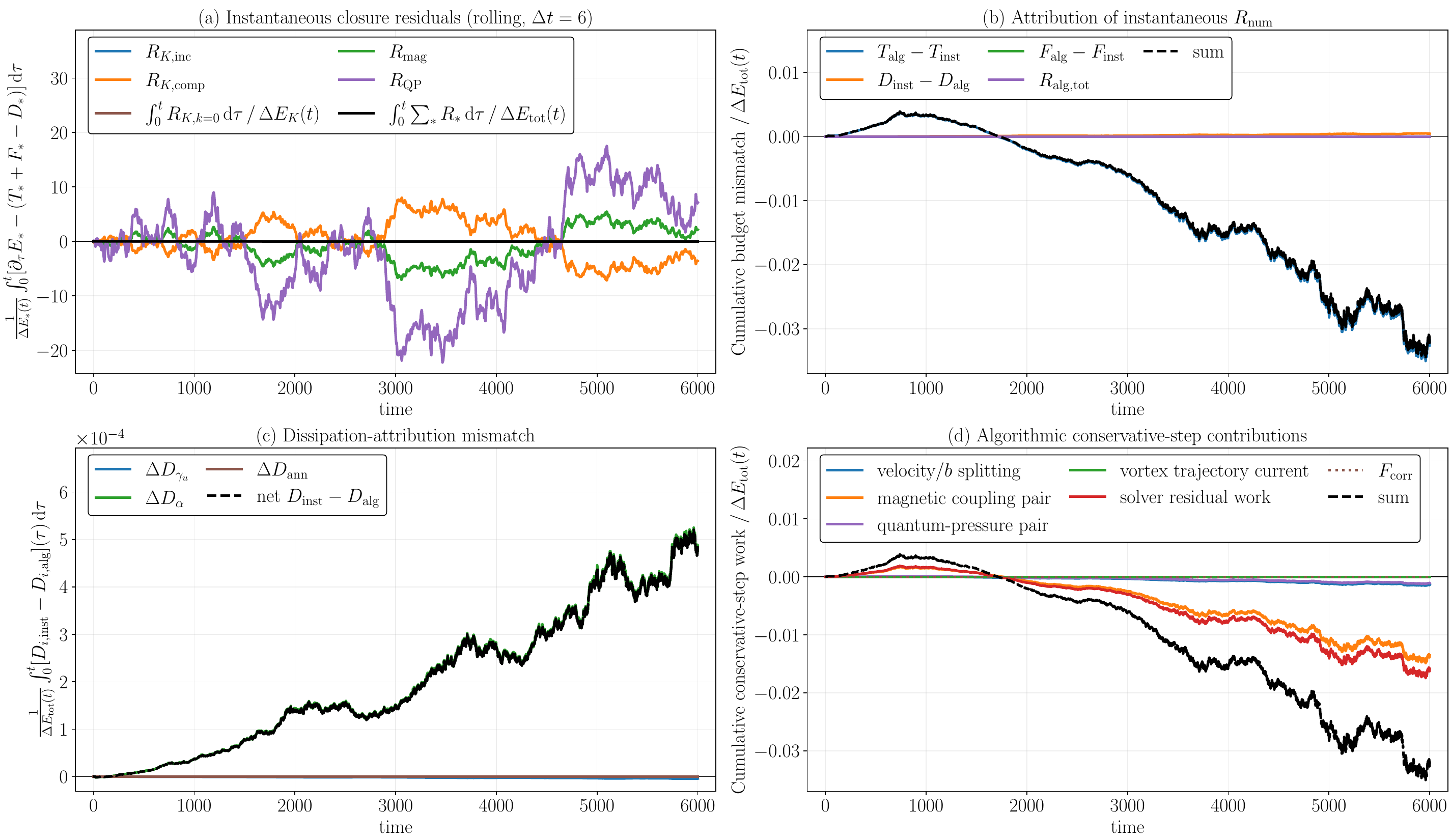}
    \caption{Cumulative energy-budget closure and its attribution. \textbf{(a)} Time-integrated instantaneous channel residuals, normalized by the corresponding energy changes $\Delta E_*(t)=E_*(t)-E_*(0)$ and smoothed with the indicated rolling window. The zero-mode kinetic residual $R_{K,k=0}$ is plotted with respect to $\Delta E_{kin}(t)$, and the total residal with respect to $\Delta E_{\mathrm{tot}}$. \textbf{(b)} Attribution of the numerical residual to differences between algorithmic and instantaneous transfer, forcing and dissipation contributions, together with the algorithmic closure residual. \textbf{(c)} Dissipation mismatch resolved into the active physical and filtering contributions. $D_\alpha$ denotes vortex mutual friction. \textbf{(d)} Conservative algorithmic step contributions, including channel exchange, vortex-trajectory current, solver-residual work and force corrections. Panels \textbf{(b-d)} are normalized by $\Delta E_{\mathrm{tot}}(t)$. ``Instantaneous'' refers to rates evaluated from the evolution equations on the spatial grid, while ``algorithmic'' refers to the energy changes produced by the finite update steps.}
    \label{fig:S3}
\end{figure*}

\begin{figure*}[!t]
   \centering
    \includegraphics[width=\linewidth]{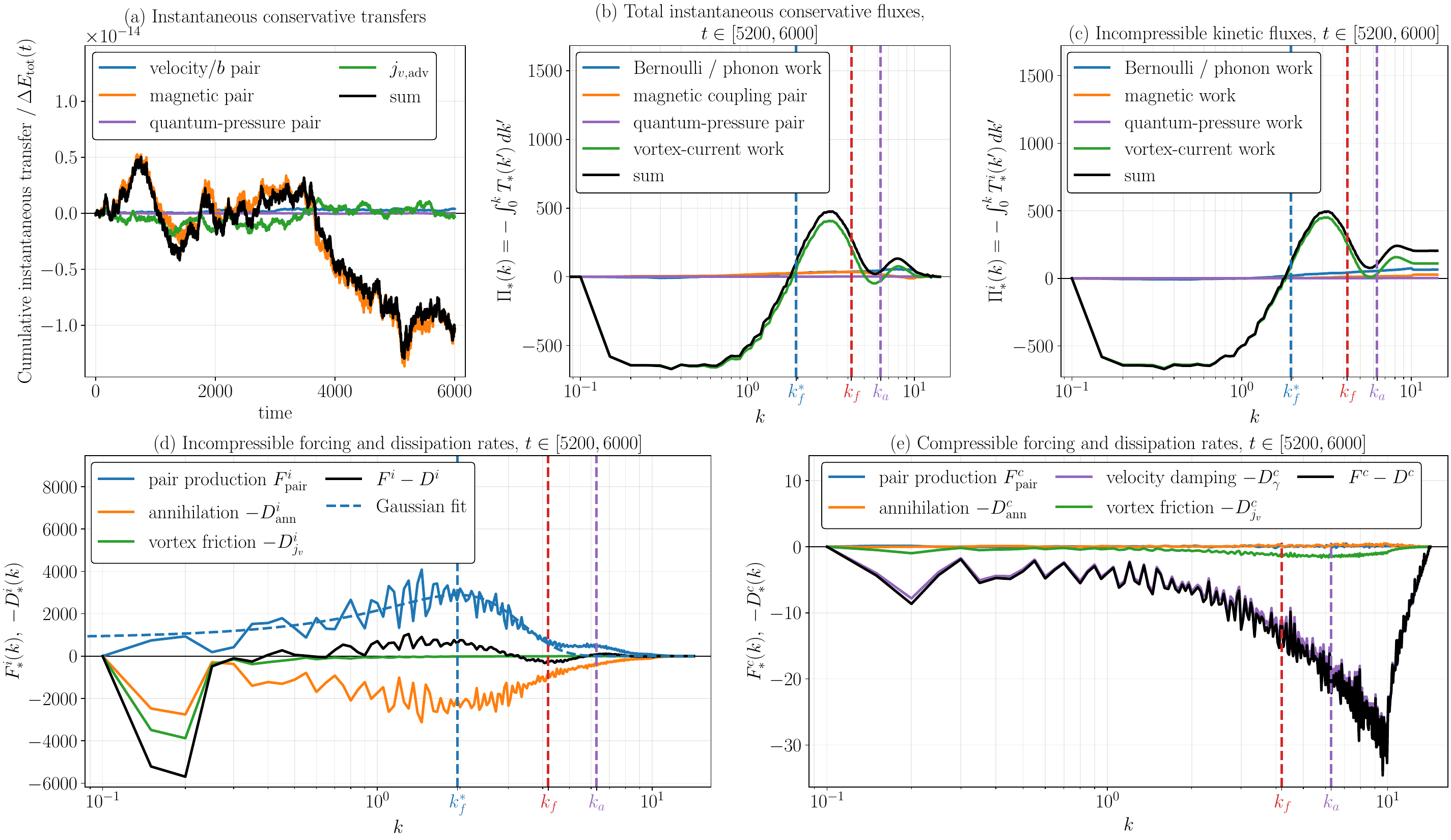}
    \caption{Conservative transfers and scale-resolved energy budgets. \textbf{(a)} Cumulative instantaneous conservative-channel transfers, including the magnetic and quantum-pressure partners, normalized by $\Delta E_{\mathrm{tot}}(t)$. \textbf{(b)} Total conservative fluxes and \textbf{(c)} incompressible kinetic-energy fluxes, decomposed into Bernoulli / phonon-work ($\mathrm{p}$), magnetic ($\mathrm{g}$), quantum-pressure ($\mathrm{Q}$) and advective vortex-current ($\mathrm{v}$) channels. Fluxes are negative cumulative shell integrals, so negative values indicate transfer toward smaller wavenumbers. \textbf{(d,e)} Signed forcing and dissipation shell rates for incompressible and compressible kinetic energy, respectively. The dashed blue Gaussian fit to vortex-pair production determines $k_f^*$ at its center position. Colored scale markers identify $k_f$, $k_f^*$, $k_\xi$ and $k_a$ where shown. Time averaging is performed over the time windows stated in the titles.}
    \label{fig:S4}
\end{figure*}

\begin{figure*}
   \centering
    \includegraphics[width=\linewidth]{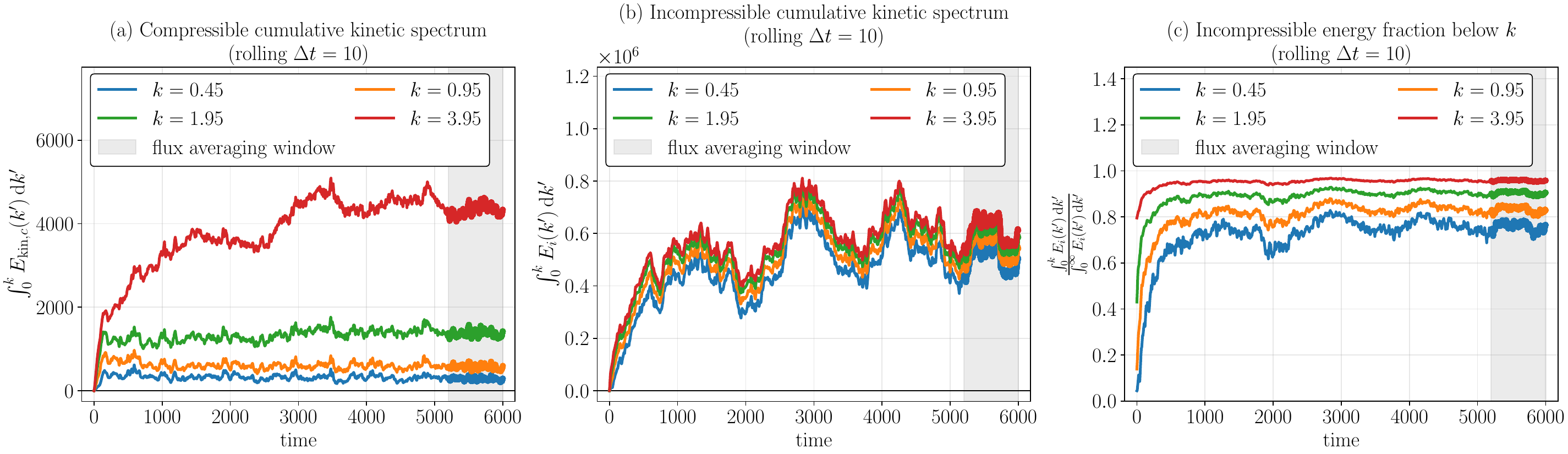}
    \caption{Cumulative kinetic spectra. \textbf{(a,b)} Compressible and incompressible kinetic energy below the four indicated wavenumbers, respectively. \textbf{(c)} Fraction of the incompressible kinetic energy below each indicated wavenumber. Panels \textbf{(a-c)} are rolling averages with the averaging windows specified in the titles and are shifted to vanish at $t=0$ after smoothing. Shading and emphasized points in \textbf{(a-c)} identify the flux-averaging interval.}
    \label{fig:S5}
\end{figure*}

\subsection{Magnetic and quantum-pressure energy fluxes}

The interaction energy and the quantum-pressure energy are
\begin{align}
  E_b &=\frac{g}{8\pi^2}\int\dd^2x\,b^2 =\frac g2\int\dd^2x\,\rho^2,\\
  E_Q &=\frac12\int\dd^2x\,|\nabla\sqrt\rho|^2.
\end{align}
Their shell spectra are therefore
\begin{align}
  E_b(k,t) &=\mathcal \int_\Omega\frac g2|\rho(\bm k,t)|^2,\\
  E_Q(k,t) &=\mathcal \int_\Omega\frac{k^2}{2}|s(\bm k,t)|^2, \qquad s=\sqrt\rho.
  \label{eq:density-energy-spectra}
\end{align}
From the continuity equation, we obtain the explicit Fourier-space rates
\begin{align}
  \mathcal F[\partial_t\rho](\bm k) &=-i\bm k\cdot\int_{\bm p}s(\bm k-\bm p)\bm u(\bm p),\\
  \mathcal F[\partial_t s](\bm k) &=-\frac{i}{2}\int_{\bm p}\int_{\bm q} s^{-1}(\bm k-\bm p)\, \bm p\cdot\left[s(\bm p-\bm q)\bm u(\bm q)\right].
  \label{eq:density-energy-fourier-rates}
\end{align}
It follows that the shell-resolved transfers into these two energy sectors are
\begin{align}
  T_b(k,t) &\equiv g\mathcal \int_\Omega\operatorname{Im}\left[ \rho^*(\bm k,t)\, \bm k\cdot\int_{\bm p}s(\bm k-\bm p)\bm u(\bm p) \right],\\
  \nonumber T_{\mathrm{Q}}(k,t)&\equiv\frac{k^2}{2}\mathcal \int_\Omega\operatorname{Im}\bigg[ s^*(\bm k,t)\int_{\bm p}\int_{\bm q} s^{-1}(\bm k-\bm p)\times \\
  &\hspace{10pt}\bm p\cdot\left(s(\bm p-\bm q)\bm u(\bm q)\right)\bigg].
  \label{eq:density-energy-transfers}
\end{align}
The corresponding cumulative fluxes are
\begin{align}
  \Pi_b(k,t)&\equiv-\int_0^k\dd k'\,T_b(k',t),\\
  \Pi_{\mathrm{Q}}(k,t)&\equiv-\int_0^k\dd k'\,T_{\mathrm{Q}}(k',t),
\end{align}
with the energy balance equation
\begin{align}  
  \partial_tE_b(k,t)+\partial_k\Pi_b(k,t)&=0,\\
  \partial_tE_Q(k,t)+\partial_k\Pi_{\mathrm{Q}}(k,t)&=0.
  \label{eq:density-energy-fluxes}
\end{align}

\subsection{Decomposition in terms of independent flux channels}

The transfer terms above describe both redistribution across scales and conversion between energy sectors~\cite{verma2022variable}. A distinction between interscale transfer and exchange between fields is also made in spectral analyses of magnetohydrodynamic turbulence~\cite{dar2001energy}. To distinguish these processes, define the full kinetic transfer $T_\chi^{\rm K}\equiv T_\chi^c+T_\chi^i$ for each channel $\chi$. For periodic boundaries, the conservative terms obey
\begin{align}
 \int_0^\infty\dd k\,T_{\mathrm{p}}^{\rm K}&=0,
 &\int_0^\infty\dd k\,T_{\mathrm{v}}^{\rm K}&=0,\\
 \int_0^\infty\dd k\,(T_{\mathrm{g}}^{\rm K}+T_b)&=0,
 &\int_0^\infty\dd k\,(T_{\mathrm{Q}}^{\rm K}+T_{\mathrm{Q}})&=0.
 \label{eq:independent-channel-cancellations}
\end{align}
The first cancellation combines the Bernoulli term with the change of the density weight in $\bm u=\sqrt\rho\,\bm v$. Together they transport kinetic energy without changing its total. The advective vortex current likewise does no net work, since its force is perpendicular to the local velocity. The remaining identities express reversible conversion between kinetic energy and, respectively, interaction and quantum-pressure energy. These cancellations hold after integration over all scales, rather than shell by shell.
Accordingly, four conservative flux channels can be grouped as
\begin{align}
 \Pi_{\mathrm{p}}&=\sum_{a=c,i}\Pi_{\mathrm{p}}^a,
 &\Pi_{\mathrm{v}}&=\sum_{a=c,i}\Pi_{\mathrm{v}}^a,\\
 \Pi_{\mathrm{g},\rm tot}&=\sum_{a=c,i}\Pi_{\mathrm{g}}^a+\Pi_b,
 &\Pi_{\mathrm{Q},\rm tot}&=\sum_{a=c,i}\Pi_{\mathrm{Q}}^a+\Pi_{\mathrm{Q}}.
 \label{eq:independent-flux-channels}
\end{align}
Their sum gives the conservative flux of the total resolved energy $E_{\rm kin}+E_b+E_Q$, and each vanishes at the upper end of the complete spectrum. A projected kinetic flux can additionally contain conversion between compressible and incompressible motion. It therefore need not describe transport within a closed energy sector. In the strict constant-density incompressible limit, the incompressible Bernoulli and density-weighting contributions vanish, and the advective vortex-current channel carries the nonlinear Euler transfer.

Mutual friction, compressible-velocity damping, and vortex creation or annihilation are external source or sink terms and are excluded from these conservative channels. In a statistically stationary state their net input balances the net dissipation, including the spectral velocity damping. Within a range where conversion, forcing and dissipation are negligible, a nearly constant incompressible flux diagnoses a cascade. The direction of energy transfer with Eq.~\eqref{eq:flux-cumulative-definition} is such that negative flux denotes transfer toward larger spatial scales.

\subsection{Numerical evaluation of the flux channels}

The numerical energy transfers are evaluated semidiscretely: Real space (the torus) is discretized on the simulation grid, while each transfer channel is obtained from the instantaneous right-hand side at a common diagnostic time. We use $\bm W=\sqrt{mb/(2\pi)}\,\bm w$, whose quadratic norm gives the kinetic energy. Since $\bm v=-\hat{\bm z}\times\bm w$, the longitudinal projection of $\bm W$ represents incompressible modes of this field, and its transverse projection represents compressible modes. For a contribution $\dot{\bm W}_\chi$ in channel $\chi$ to the instantaneous evolution, the rate in the shell $s$ is
\begin{equation}
 T_{\chi,h}^A(k_s)=\frac{h^2}{N^2\Delta k}
 \sum_{\bm k\in s}\operatorname{Re}\left[
   (P^A\widetilde{\bm W}_{\bm k})^*\cdot
   \widetilde{\dot{\bm W}}_{\chi,\bm k}\right],
 \hspace{5pt} A=L,T.
 \label{eq:semidiscrete-shell-work}
\end{equation}
Here $P^A$ with $A=L,T$ denotes the projector. The full kinetic rate uses the identity projector $P^A = 1$, including $\bm k=0$, while the projected rates exclude this mode. Shells have width $\Delta k=0.05$, with no division by the number of modes in a shell. The ordinary energy spectra are saved with shell width $0.1$ and displayed in bins of width $0.4$. Displayed transfer rates use overlapping intervals of width $0.1$, shifted by $0.05$; cumulative fluxes are sampled from the fine-grid cumulative sums at the upper ends of those intervals. All derivatives use the simulation's spectral operators and nonlinear products are formed on the original grid. Velocity-force contributions enter as $\dot{\bm W}_\chi=\sqrt{mb/(2\pi)}\,\dot{\bm w}_\chi$, whereas the density-weighting term is $\bm W\dot b/(2b)$, with $\dot b=-\curl(b\bm w)$ for conservative transport.

Quantum-pressure work uses the variational derivative of the actual grid energy in Eq.~\eqref{eq:EQ},
\begin{equation}
 Q_h=2\pi\frac{\delta E_{Q,h}}{\delta b}
 =\frac{1}{8m}\left[-\frac{|\nabla_h b|^2}{b^2}
       -2\div_h\left(\frac{\nabla_h b}{b}\right)\right].
 \label{eq:semidiscrete-quantum-pressure}
\end{equation}
The interaction and quantum-energy shell rates are obtained by differentiating their quadratic grid fields $\sqrt g\,b/(2\pi)$ and $\nabla_hb/\sqrt{8\pi m b}$ along the same $\dot b$. Thus the quantum-energy shell decomposition follows the saved grid Hamiltonian. Its exchange with kinetic energy cancels after summing all modes, to roundoff.

The instantaneous advective vortex current deposits $q_a\bm v_{s,a}$ using the same resolved spectral cloud and density-weighted interpolation as the particle dynamics. Mutual-friction work uses the difference between the full and advective deposited currents. These currents are evaluated at the diagnostic positions, rather than integrated along a finite-step trajectory. The matched interpolation and deposition make the full-spectrum advective work vanish to roundoff. Velocity damping is evaluated from its instantaneous spectral generator, using $\dot{\bm w}_\gamma=-\gamma\bm w^T$ in the shell-work formula, consistently with Eq.~\eqref{eq:flux-velocity-damping}.

The conservative curves combine the four channels in Eq.~\eqref{eq:independent-flux-channels}. The total-energy curves include the interaction and quantum-energy partners, whereas each projected kinetic curve includes only its kinetic contribution. For every conservative channel, the plotted flux is the negative cumulative shell sum,
\begin{equation}
 \Pi_{\chi,h}^A(k_s)=-\Delta k\sum_{r\leq s}T_{\chi,h}^A(k_r).
\end{equation}
Forcing and dissipation are instead displayed as signed shell rates, with losses entering negatively. Pair-production and annihilation rates are obtained from a comparison of the actual kinetic spectral changes before and after an event occurs, accumulated over the diagnostic interval and divided by its duration. Their signs are retained.

We compute a numerical residual defined by the difference between the measured endpoint spectral-energy rate and the averaged semidiscrete balance. It includes temporal discretization, finite solve tolerances, sampling and splitting effects. This residual is a numerical artefact and is not assigned to a conservative cascade channel.

\Cref{fig:S3} separates this cumulative closure residual into instantaneous channel balances, mismatches with the finite-step algorithmic contributions, and the resolved conservative-step work. Here $\Delta E_*(t)=E_*(t)-E_*(0)$ denotes the energy change used for normalization. 
\Cref{fig:S4} shows the conservative transfer decomposition and the wavenumber-resolved fluxes and forcing/dissipation rates. The instantaneous rates are evaluated from the evolution equations at a common state, whereas algorithmic contributions are measured from the finite discrete update steps. A Gaussian fit to the pair-production shell rate defines the effective forcing wavenumber $k_f^*$.
\Cref{fig:S5} shows the time evolution of the compressible and incompressible kinetic energies integrated up to selected wavenumbers, together with the corresponding fraction of the incompressible kinetic energy below certain wavenumbers relative to the total incompressible kinetic energy.

\begin{figure*}[!t]
   \centering
    \includegraphics[width=\linewidth]{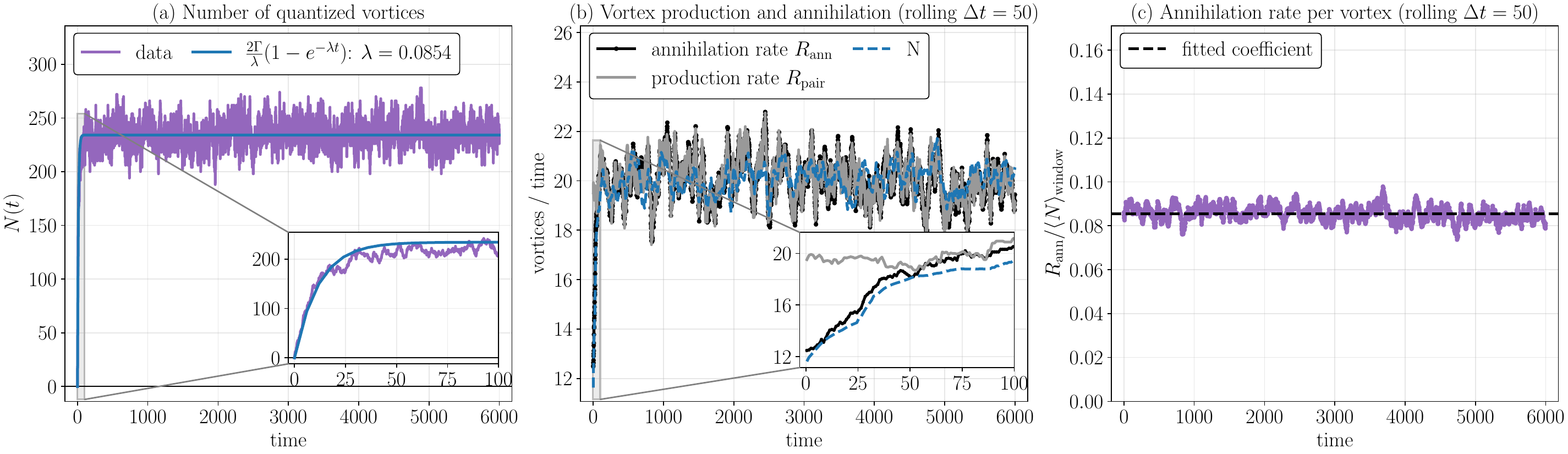}
    \caption{Vortex-population dynamics and annihilation-rate normalization. \textbf{(a)} Total number of quantized vortices as a function of time and fit $N(t)=2\Gamma/\lambda\, (1-e^{-\lambda t})$, with a fit parameters $\lambda$. The steady state vortex number is given by $N_0 =2\Gamma/\lambda$ with $\Gamma$ being the vortex pair production rate. \textbf{(b)} Vortex production and annihilation rates, $R_{\mathrm{pair}}$ and $R_{\mathrm{ann}}$, compared to the window-mean vortex population.  \textbf{(c)} Annihilation rate divided by the window-mean population, $R_{\mathrm{ann}}/\langle N\rangle_{\mathrm{window}}$. The horizontal dashed line is the fitted coefficient of the linear population dependence. Panels \textbf{(b,c)} use the rolling windows indicated in their titles. Insets show the early-time evolution.}
    \label{fig:S6}
\end{figure*}

\section{Vortex clustering in turbulent state}\label{app:clustering}
\subsection{Vortex classification}

Each saved vortex configuration is partitioned into four disjoint classes: dipoles, positive clusters, negative clusters and isolated vortices. We use a recursive classification based on nearest-neighbor criteria employed in earlier studies~\cite{reeves2013inverse,groszek2018vortex,valani2018einstein}. All distances use the minimum-image convention on the periodic domain,
\begin{equation}
 d_{ab}^2=\sum_{\mu=x,y}\min\bigl(|X_{a,\mu}-X_{b,\mu}|,
                   L_\mu-|X_{a,\mu}-X_{b,\mu}|\bigr)^2.
\end{equation}
First, opposite-sign vortices that are mutual nearest neighbors are identified as dipoles and removed from the active list. Nearest neighbors are recomputed and this procedure is repeated until no further dipoles remain. Note that the classification does not remove particles from the simulation and uses no fixed dipole-distance threshold.

For the remaining vortices, let $d_a^{\rm opp}$ be the distance of vortex $a$ to the nearest active vortex of opposite sign, or infinity if none exists. Two like-signed vortices are connected when
\begin{equation}
 q_a=q_b,\qquad d_{ab}<\min(d_a^{\rm opp},d_b^{\rm opp}).
 \label{eq:vortex-cluster-link}
\end{equation}
Connected components with at least two members define same-sign clusters. These components are assigned their vortex sign and removed from the active list. This procedure is continued and the graph recomputed until no further clusters are found. All remaining vortices are classified as isolated. Thus a cluster is a connected component and need not have every pair of members directly linked. The displayed connections within each cluster form a minimum spanning tree using the periodic distances. Their purpose is to visualize the clusters in real space.

\subsection{Vortex distribution}

To obtain a visualization of vortex clustering, we compute the spatial vortex distribution averaged over space and time and decompose the distribution in terms of the different vortex classes. 
The spatial distribution map averages the classified vortex positions over a selected time window. We divide the periodic box into rectangular bins spanning three simulation cells in each direction and classify every saved configuration in the window independently. For class $\lambda$, the density in bin $B$ is
\begin{equation}
 \bar n_\lambda(B)=\frac{1}{M\abs{B}}
    \sum_{\ell=1}^{M}\sum_{a\in\lambda(t_\ell)}
       \bm 1_B(\bm X_a(t_\ell)),
 \label{eq:vortex-class-distribution}
\end{equation}
where $M$ is the number of selected temporal frames and $\abs{B}$ is the bin area. This amounts to an equal-frame average of number density, rather than a probability density normalized separately for each class.

The four densities are combined in one image. Orange denotes isolated vortices, green dipoles, blue positive clusters and red negative clusters. Each bin's hue is the density-fraction-weighted mixture of these colors. Its contrast against a light background increases with the total density $\bar n=\sum_\lambda\bar n_\lambda$, using $\min[\bar n/(0.60\bar n_{\max}),1]^{0.60}$. The image therefore displays both local class composition and vortex concentration, with saturated contrast at high density.

\subsection{Vortex correlations}

To quantify the amount of vortex clustering, we compute vortex sign correlations~\cite{white2012creation,reeves2013inverse,yang2025quantum,jha2025energy} for a certain number of vortex neighbors. 
Local sign correlations are computed from all vortices before classification. Let $N$ be the instantaneous vortex number and $a_j$ the $j$th nearest neighbor of vortex $a$, ordered by periodic distance. The plotted measure in the main text is
\begin{equation}
 C_p=\frac{1}{p N}\sum_{a=1}^{N}\sum_{j=1}^{p}q_aq_{a_j}
 \label{eq:vortex-sign-correlations}
\end{equation}
for $p=4$.
Positive correlations indicate a preference for like-signed neighbors, while negative correlations indicate opposite-sign pairing. The reference for randomly permuted signs at fixed total charge is
\begin{equation}
 C_{\rm rand}=\frac{(\sum_aq_a)^2-N}{N(N-1)}
             =-\frac{1}{N-1}
 \label{eq:vortex-random-sign-reference}
\end{equation}
for a neutral configuration. The finite-population reference is therefore slightly negative.

The accompanying cluster fractions~\cite{groszek2018vortex} quantify how many vortices participate in collective structures. If the same-sign components defined by Eq.~\eqref{eq:vortex-cluster-link} have sizes $s_1\geq s_2\geq\cdots\geq2$, we report
\begin{equation}
 f_{\rm cl}=\frac{\sum_\nu s_\nu}{N},\qquad
 f_4=\frac{\sum_{\nu=1}^{\min(4,N_{\rm cl})}s_\nu}{N}.
\end{equation}
Both fractions use the total vortex number, including dipoles and isolated vortices, as denominator. The four largest components are selected jointly from both signs. Thus $f_{\rm cl}$ measures overall cluster participation, whereas $f_4$ measures the concentration of vortices in the few largest clusters. Hence, $f_4$ is a proxy for how many vortices participate in single clusters, i.e. how larger the vortex clusters are. Configurations are sampled at approximately unit time spacing, including the final saved frame, and all correlation and fraction curves are smoothed with a centered rolling mean of width 50.

\begin{figure*}[!t]
    \centering
    \includegraphics[width=\linewidth]{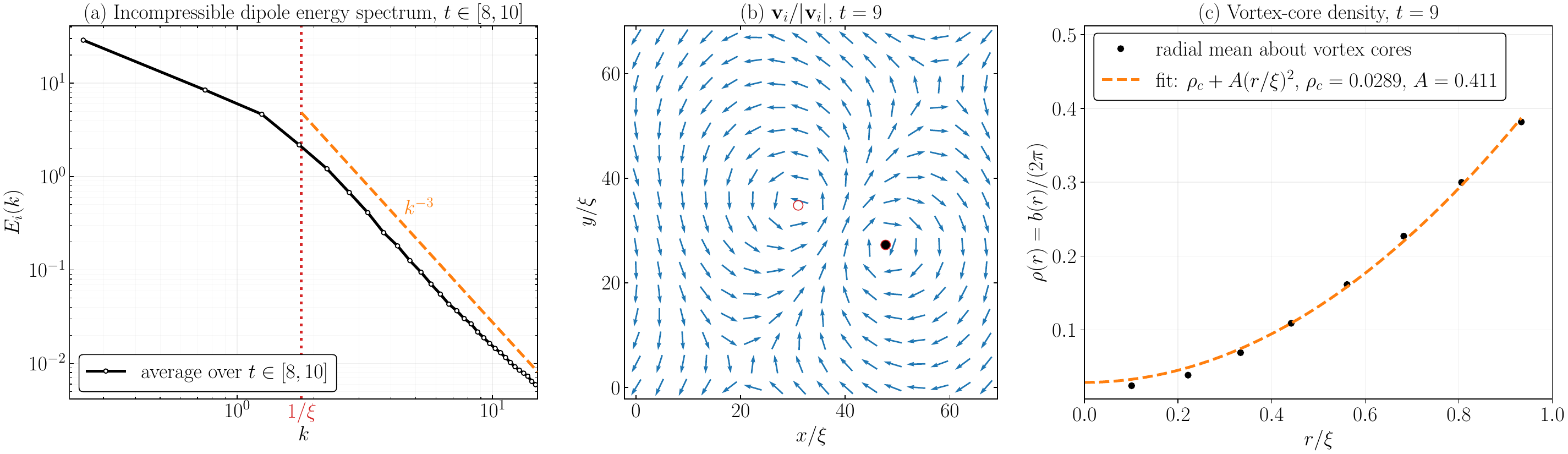}
    \caption{\textbf{(a)} Incompressible kinetic energy spectrum of a single vortex-antivortex pair at $g=1$, $b_0 = 10.0$ ($g b_0/(2\pi) = 1.6$). At scales $k\xi \gg 1$, the spectrum follows a $k^{-3}$ power law scaling, as indicated by the orange reference line. \textbf{(b)} Snapshot of the direction of the incompressible velocity field of the dipole at $t=9.0$. \textbf{(c)} Density profile of both vortices averaged in radial shells close to the vortex core with $r < \xi$ as black points. Quadratic fit in orange.}
    \label{fig:S7}
\end{figure*}

\section{Vortex number steady state}\label{app:vortex_steady_state}

In this appendix, we describe a simple model for the evolution of the vortex number, as shown in \cref{fig:S6}. In our dynamics, vortex production is implemented through a Poisson process with rate $\Gamma$ and annihilated whenever they are close enough with separation below a threshold. Since $\Gamma$ measures the number of pair-creation events over time, and each creation process increases the number of vortices by two, we obtain the following candidate differential equation,
\begin{equation}
    \frac{\dd N}{\dd t}= 2\Gamma -R_{\textrm{ann}}(t)
\end{equation}
where $R_{\textrm{ann}}(t)$ is an annihilation rate. The vortex annihilation rate is likely to depend on some power of the number of vortices in the superfluid; that is, $R_{\textrm{ann}}(t)\propto N^p$. Upon measuring the annihilation rate (as shown in Fig.~~\ref{fig:S6}(b,c)), we find that the annihilation rate scales linearly with the number of vortices, thus motivating the differential equation,
\begin{equation}
    \frac{\dd N}{\dd t}= 2\Gamma -\lambda N
\end{equation}
for a annihilation parameter $\lambda$, which we obtain by fitting to this data. This equation is solved by an exponential growth to a steady-state---that is,
\begin{equation}
    N(t)=\frac{2\Gamma}{\lambda}(1-e^{-\lambda t})
\end{equation}
The steady-state vortex number is obtained by $2\Gamma/\lambda$. Fitting to the data yields $\lambda\approx 0.085$.

\section{Kinetic energy scaling at scales within the vortex core}\label{app:dipole_scaling}

The $k^{-3}$ scaling of the incompressible energy spectrum for $k\gg \xi^{-1}$ can be understood by a simple scaling argument. In the neighborhood of a vortex (taken to be at the origin), $\rho(r) = \vert \psi(r) \vert^2 \sim r^2$. To see this, consider a point vortex solution of charge $n$ of the GP equation (Eq.~\eqref{GP_equation}), which amounts to a solution of the form (in polar coordinates $(r,\theta)$),
\begin{equation}
    \psi_n(r,\theta)=f(r)e^{\textrm{i} n\theta}.
\end{equation}
where $f(r)$ is an unknown function that sets the density profile of the vortex, and $n$ is the associated winding number. Inserting this into the GP equation yields the following differential equation for $f(r)$, 
\begin{equation}\label{f_ode}
    -\frac{\hbar^2}{2m}\left(\partial_r^2+\frac{1}{r}\partial_r-\frac{n^2}{r^2}\right)f(r)+gf^3(r)=\mu f(r)
\end{equation}
This equation may be non-dimensionalized as per the transformations,
\begin{equation}
    x=r/\xi\;\;\; \textrm{and}\;\;\; f(r)=\Phi(x)\sqrt{\rho_0},
\end{equation}
noting that the constant (average) density solution of Eq.~\eqref{GP_equation} is $\rho_0=\mu/g$. Then, we find a differential equation for $\Phi(x)$ given by,
\begin{equation}\label{vortex_solution_ode}
    \partial_x^2\Phi(x)+\frac{1}{x}\partial_x\Phi(x)-\frac{n^2}{x^2}\Phi(x)+\Phi(x)-\Phi^3(x)=0.
\end{equation}
This is a non-linear (cubic) second-order differential equation with singular coefficients (as $x\rightarrow 0$). While this equation cannot be solved exactly in terms of special functions, one can study its asymptotic properties of the solution $\Phi(x)$ of the differential equation in Eq.~\eqref{vortex_solution_ode}:  
\begin{enumerate}
    \item $x\rightarrow 0$:  Note that the differential equation in Eq.~\eqref{vortex_solution_ode} has a regular singular point at $x=0$. Therefore, one can use the Frobenius method, as per which we find,
    \begin{equation}\label{small_x}
        \Phi(x)\sim x^n\left(1-\frac{1}{4(n+1)}x^2+O(x^4)\right)\;\;\; \textrm{at small }x,
    \end{equation}
    where $\sim$ indicates proportionality upto an unknown constant. 
    \item $x\rightarrow \infty$: We can take $\Phi(x)=1-\eta(x)$ in this limit, and then obtain a differential equation from Eq.~\eqref{vortex_solution_ode} keeping only linear order in $\eta$. Then, we obtain, 
    \begin{equation}
           \left( \partial_x^2 + \frac{1}{x}\partial_x -\frac{n^2}{x^2}-2\right)\eta(x)=-\frac{n^2}{x^2}.
    \end{equation}
    The homogeneous solution of this equation is a modified Bessel function of the second kind. The particular solution dominates the behavior at large $x$, and we obtain asymptotically,
    \begin{equation}\label{large_x}
        \Phi(x)=1-\frac{n^2/2}{x^2}+O(x^{-4}) \;\;\; \textrm{at large } x.
    \end{equation}
    This implies a polynomial decay away from the saturation at $x\rightarrow \infty$. 
\end{enumerate}
This yields that $\rho\propto \Phi^2(r)\sim r^2$ at small $r$ for vortex charges $n=\pm 1$. We verify this in our simulations in \cref{fig:S7}(c). 

Going back to the argument for a $k^{-3}$ scaling, we note that the velocity is purely azimuthal with $v \sim 1/r$, which implies the flow is locally incompressible. As a consequence, at short distances $r < \xi$, $\int \dd^2 r^\prime \, \rho(r')v^2(r') \sim r^2$. Equating this to a kinetic energy spectrum, $r^2 \sim \int \, \dd k \, E_i(k)$, we can immediately deduce that $E_i(k) \sim k^{-3}$. Note that since $E_i(k)$ in this regime is ascribed to how density depletes at a single vortex, it does \emph{not} represent a kinetic energy cascade. We show such a scaling in \cref{fig:S7}(a) obtained from our simulations, with the corresponding incompressible velocity field of the dipole visualized in \cref{fig:S7}(b).

\makeatletter
\renewcommand{\bibsection}{%
  \par
  \begingroup
    \baselineskip26\p@
    \bib@device{\columnwidth}{0.25\columnwidth}%
  \endgroup
  \nobreak\@nobreaktrue
  \addvspace{19\p@}%
  \par
}
\makeatother

\end{document}